\documentclass[12pt]{iopart}

\usepackage{iopams}  
\usepackage{graphicx}

\def\bea{\begin{eqnarray}}
\def\eea{\end{eqnarray}}

\begin{document}


\title{Applicability of Monte Carlo Glauber models to relativistic heavy ion collision data}

\author{R L Ray and M S Daugherity}

\address{Department of Physics, The University of Texas at Austin, Austin, Texas 78712 USA}
\ead{ray@physics.utexas.edu}

\begin{abstract}
The accuracy of Monte Carlo Glauber model descriptions of minimum-bias multiplicity frequency distributions is evaluated using data from the Relativistic Heavy Ion Collider (RHIC) within the context of a sensitive, power-law representation introduced previously by Trainor and Prindle (TP). Uncertainties in the Glauber model input and in the mid-rapidity multiplicity frequency distribution data are reviewed and estimated using the TP centrality methodology. The resulting errors in model-dependent geometrical quantities used to characterize heavy ion collisions ({\em i.e.} impact parameter, number of nucleon participants $N_{part}$, number of binary interactions $N_{bin}$, and average number of binary collisions per incident participant nucleon $\nu$) are presented for minimum-bias Au-Au collisions at $\sqrt{s_{NN}}$ = 20, 62, 130 and 200~GeV and Cu-Cu collisions at $\sqrt{s_{NN}}$ = 62 and 200~GeV. Considerable improvement in the accuracy of collision geometry quantities is obtained compared to previous Monte Carlo Glauber model studies, confirming the TP conclusions. The present analysis provides a comprehensive list of the sources of uncertainty and the resulting errors in the above geometrical collision quantities as functions of centrality. The capability of energy deposition data from trigger detectors to enable further improvements in the accuracy of collision geometry quantities is also discussed.
\end{abstract}

\submitto{\JPG}

\maketitle

\section{Introduction}
\label{Sec:Intro}

Observables in relativistic heavy ion experiments are often reported as functions of measured total inelastic cross section fraction (centrality) and are related to the initial collision geometry using Monte Carlo Glauber (MCG) or optical Glauber models~\cite{starmc,glauber-rev,glauber}. An example is charged particle multiplicity $N_{ch}$ versus number of participating nucleons $N_{part}$. It is therefore imperative that the level of accuracy of the Glauber models be understood and that relevant experimental information be used to estimate collision centrality. For instance, previous studies~\cite{starmc,glauber-rev,starspec} concluded that the uncertainties in relating peripheral collision data to the corresponding initial collision geometry were very large, thus contributing to the omission of those data from publications~\cite{starmc}. In another example, studies of high-$p_t$ suppression of charged particle production for peripheral to mid-central collisions are presently limited by large uncertainties in the estimated number of binary collisions which is used as a scaling variable~\cite{highpt}.

In this work we re-examine the accuracy of Monte Carlo Glauber model descriptions of minimum-bias multiplicity frequency distribution data from RHIC using updated density and multipliciity production model parameters and a sensitive power-law inspired representation introduced by Trainor and Prindle (TP)~\cite{powerlaw}. After demonstrating the accuracy of our Monte Carlo Glauber model in describing RHIC data we then use the power-law analysis method of Ref.~\cite{powerlaw} to estimate the uncertainties in the mapping relationships between initial stage collision geometry quantities and centrality for heavy ion collision systems relevant to the RHIC program. We find that previous uncertainties~\cite{starmc,glauber-rev,starspec} in centrality measures are too pessimistic.

The centrality analysis method of Trainor and Prindle~\cite{powerlaw}
exploits the
approximate power-law dependence of the multiplicity frequency
distribution data $dN_{evt}/dN_{ch}$ ($N_{evt}$ is the number of triggered
events) for minimum-bias~\footnote{A minimum-bias trigger
typically refers to the detection of forward going
spectator fragments from both colliding nuclei plus
a minimal requirement for particle
production transverse to the beam direction.
Each of the four RHIC experiments utilize a common lowest level trigger
detector system based on calorimetric detection of neutrons at zero degree
scattering angle using two zero-degree calorimeters (ZDC) placed
symmetrically upstream and downstream from the beam-beam intersection
region. The minimum-bias trigger systems for the STAR, PHENIX, PHOBOS and
BRAHMS experiments are described in Nucl. Instrum. Meth. A {\bf 499}, Nos. 2-3
(2003).} collisions and uses
proton-proton (p-p) multiplicity production data to constrain the peripheral
collision end-point of the $dN_{evt}/dN_{ch}$ distribution.
Their analysis demonstrates that uncertainties in centrality determination and
MCG geometry measures can be significantly reduced, particularly in
the peripheral region.

In this paper MCG results and errors for centrality bin average quantities
$\langle b \rangle$ (mean impact parameter), $\langle N_{part} \rangle$,
$\langle N_{bin} \rangle$, and $\nu = \langle N_{bin} \rangle/(\langle N_{part} \rangle /2)$ ~\cite{tomnu}
are presented for minimum-bias Au-Au
collisions at $\sqrt{s_{NN}}$ = 20, 62, 130 and 200~GeV and for Cu-Cu at
62 and 200~GeV.  In this study centrality is based on
charged particle production at midrapidity. Other centrality definitions
appear in the literature but will not be considered here.
However, the MCG model
and analysis methods presented here and in Ref.~\cite{powerlaw} can be readily
extended to any practical centrality determination method.

The MCG model and updated input parameters are discussed in Sec.~\ref{Sec:model}.
The accuracy of the model for describing
$dN_{evt}/dN_{ch}$ data from RHIC is demonstrated in Sec.~\ref{Sec:datafit}.
MCG predictions and errors are presented and discussed in
Sec.~\ref{Sec:results}. Comparison of the MCG results and analytic parametrizations from Ref.~\cite{powerlaw} are presented in Sec.~\ref{Sec:analytic}. In Sec.~\ref{Sec:discussion} we demonstrate via simulations 
how energy deposition data from transverse
particle production, which are generally available at the trigger level from the RHIC experiments~\cite{dunlop},
can be used to further reduce
the impact of background contamination and to mitigate the effects of event loss due to
trigger and collision vertex
finding inefficiencies.
A summary and conclusions are presented
in Sec.~\ref{Sec:summary}. Computational details are contained in three
appendixes at the end.

\section{Monte Carlo Glauber Model}
\label{Sec:model}

The Monte Carlo Glauber collision model used here is based on a standard
set of assumptions~\cite{glauber} which are appropriate for ultra-relativistic heavy ion
collisions.  These assumptions include the characterization of the collision
in terms of a classical impact parameter $(b)$, straight-line propagation of
each incident nucleon through the oncoming nucleus, and a fixed transverse
interaction range determined by the nucleon-nucleon (N-N) total inelastic cross
section ($\sigma_{inel}$) in free space.
The impact parameter was selected at random and the positions of the nucleons
relative to the geometrical centers of the colliding nuclei were randomly
distributed according to a spherical density $\rho(r)$. Center-of-mass constraints were not imposed on the nucleon
positions. Nucleon pairs in the colliding nuclei were assumed to interact hadronically
if their relative impact parameter was $\leq \sqrt{\sigma_{inel}/\pi}$.

Charged hadron multiplicity was assigned using the phenomenological
two-component model of Kharzeev and Nardi~\cite{kn} for ``soft'' plus
``hard'' particle production processes where the mean number of charged
hadrons in the acceptance ($\bar{N}_{ch}$) per unit pseudorapidity ($\eta$)
was computed according to
\bea
\frac{d\bar{N}_{ch}}{d\eta} & = & (1-x)n_{pp}\frac{N_{part}}{2} +
x n_{pp} N_{bin}.
\label{Eq1}
\eea
Parameters $n_{pp} \equiv d\bar{N}_{ch}(pp)/d\eta$ and $x$
depend on collision energy. 
Event-wise multiplicities were obtained by sampling the Gaussian
distribution~\cite{kn}
\bea
{\cal P}(N_{ch},\bar{N}_{ch}) & = & \frac{1}{\sqrt{2\pi a \bar{N}_{ch}}}
\exp \left( -\frac{(N_{ch} - \bar{N}_{ch})^2}{2a\bar{N}_{ch}} \right)
\label{Eq2}
\eea
where $a$ is a multiplicity fluctuation width parameter. Alternate distributions
({\em e.g.} Poisson, negative binomial~\cite{nbd}) produce quantitative effects on the
multiplicity frequency distribution for very peripheral collisions but do not
affect the present estimates of systematic errors.
$N_{bin}$ and $N_{ch}$ were both required to be $\geq 1$ in order for the simulated
collision to be used in the analysis.
Centrality bins were defined using the multiplicity frequency distribution
$dN_{evt}/dN_{ch}$. Centrality bin average quantities, $\langle b \rangle$, $\langle N_{part} \rangle$,
$\langle N_{bin} \rangle$ and
$\nu \equiv 2 \langle N_{bin} \rangle / \langle N_{part} \rangle$,
were calculated using the events within each bin. The acceptance for this
study was $|\eta| \leq 0.5$ and full $2\pi$ in azimuth.

\subsection{Matter Densities}
\label{Subsec:matterdensity}

Monte Carlo Glauber simulations require the distribution of the centers of the
nucleons in the nuclear ground state, $\rho_{pt,m}(r)$, the point-matter density.
For $^{63}$Cu and $^{197}$Au these
were estimated using the measured charge densities~\cite{dejager}
and the Hartree-Fock calculations of Negele using the Density
Matrix Expansion (DME) framework~\cite{dme}
for the neutron - proton density differences.
The charge densities for $^{63}$Cu and $^{197}$Au were represented by a
Woods-Saxon distribution,
\bea
\rho(r) & = & \rho_0 \{ 1 + \exp [(r-c)/z] \}^{-1},
\label{Eq3}
\eea
where the radius and diffuseness parameters are listed in Table~\ref{TableI}.
\begin{table}[h]
\caption{\label{TableI}
Charge and point matter density Woods-Saxon parameters for $^{63}$Cu and
$^{197}$Au in fm.}
\begin{indented}
\item[]\begin{tabular}{@{}cllll}
\br
 Density & \multicolumn{2}{c}{$^{63}$Cu} & \multicolumn{2}{c}{$^{197}$Au} \\
 Parameter &  Empirical$^{\rm a}$
 & DME$^{\rm b}$  & Empirical$^{\rm a}$ & DME$^{\rm b}$ \\
\mr 
$c_{chrg}$ & 4.214$\pm$0.026  & 4.232  & 6.38$\pm$0.06  & 6.443 \\
$z_{chrg}$ & 0.586$\pm$0.018  &        & 0.535$\pm$0.027 &      \\
$\langle r^2_{chrg} \rangle ^{1/2}$ & 3.925$\pm$0.022 & 3.899 & 5.33$\pm$0.05 & 5.423 \\
\mr 
$c_{pt,m}$ & 4.195$\pm$0.085  & 4.213  & 6.43$\pm$0.10 & 6.495 \\
$z_{pt,m}$ & 0.581$\pm$0.031  &        & 0.568$\pm$0.047 &      \\
$\langle r^2_{pt,m} \rangle^{1/2}$ & 3.901 & 3.875 & 5.41 & 5.502 \\
\br 
\end{tabular}
\item[] $^{\rm a}$ Charge density results based on electron scattering
analysis~\cite{dejager}; estimates of point matter densities as discussed in the text.
\item[] $^{\rm b}$ Density Matrix Expansion predictions~\cite{dme}.
\end{indented}
\end{table}
The
point matter densities assumed for the present analysis are spherically
symmetric with a Woods-Saxon radial distribution where the half-density
and rms radii were estimated by adding the Hartree-Fock DME point matter --
charge distribution differences to the measured charge density radii
where,
\bea
c_{pt,m} & = & c_{chrg} + \left[c_{pt,m} - c_{chrg} \right]_{\rm DME}, \nonumber \\
\langle r^2_{pt,m} \rangle ^{1/2} & = & \langle r^2_{chrg} \rangle ^{1/2}
+ \left[ \langle r^2_{pt,m} \rangle ^{1/2} - \langle r^2_{chrg} \rangle ^{1/2}
\right]_{\rm DME},
\label{Eq4}
\eea
where $c_{chrg}$ and $\langle r^2_{chrg} \rangle ^{1/2}$ denote the measured
radii from electron scattering while the quantities in the square brackets
are the DME predictions. The diffusivity parameter $z_{pt,m}$ was obtained
from the quantities in Eq.~(\ref{Eq4}) and the approximate relation,
$\langle r^2\rangle \cong \frac{3}{5} c^2 [1 + \frac{7}{3} (\pi z/c)^2 ]$~\cite{velicia}.
The nominal radius and diffusivity parameters assumed here for Au and Cu
are listed in Table~\ref{TableI}.

The uncertainties in $c_{pt,m}$ and $z_{pt,m}$ were obtained by summing
the independent errors in the proton~\cite{dejager} and neutron point matter densities in
quadrature.  The latter
is constrained by
theoretical and experimental information about the neutron - proton
density difference.  Theoretical nuclear structure model predictions
for the neutron$-$proton rms radii differences for isotopes in the Cu and Au mass range agree to
within about $\pm 0.06$~fm~\cite{cls}.  In general the theoretical
predictions agree with analyses of medium energy proton-nucleus elastic
scattering data~\cite{rayden} which are typically uncertain by about
$\pm 0.07$~fm for isotopes in the Cu and Au mass range.
The uncertainty in the neutron density rms radii relative to the proton
density was therefore assumed to be $\pm\sqrt{0.06^2 + 0.07^2}{\rm ~fm}
\approx \pm 0.09$~fm and the corresponding uncertainty in the matter rms radii
was $\pm(N/A)0.09$~fm, where $N,A$ are the
number of neutrons and nucleons in the isotope. Theoretical
contributions to the errors in $c_{pt,m}$ and $z_{pt,m}$ were conservatively
estimated by requiring each to independently account for the $\pm$(N/A)0.09~fm
error.  The latter theoretical errors were combined in quadrature with
the corresponding errors for $c_{chrg}$ and $z_{chrg}$ from analysis of
electron scattering data to obtain the final errors listed in
Table~\ref{TableI}.

\subsection{N-N Inelastic Cross Section}
\label{Subsec:cross-section}

The N-N total inelastic cross sections used here were based on total cross
section measurements for p-p collisions ($\pm$1~mb uncertainty) and
elastic total cross sections for p-p and p-$\bar{\rm p}$
($\pm$0.5~mb error)~\cite{pdg}.
Proton-neutron total cross section data are not available in the energy
range studied here.
The results for energies $\sqrt{s}$ = 20, 62, 130 and 200~GeV are respectively
33, 35.3, 38.7 and 41.7~mb, each being uncertain by $\pm$~1.1~mb.

It is possible that the effective interaction cross section between colliding
nucleons inside a nucleus differs from that in free
space (density dependence)
or that the strength and range of the effective
N-N interaction changes with each successive collision
as in the ``used'' nucleon
scenario~\cite{signneff} or the ``strict'' participant scaling
model~\cite{powerlaw} (limiting case of the used-nucleon model in which nucleons
interact only once).  The study of density dependent 
effects is well beyond the scope
of the present analysis.

\subsection{Two-Component Multiplicity Production Model}
\label{Subsec:twocomponent}

Parameters $n_{pp} = d\bar{N}_{ch}(pp)/d\eta|_{\eta=0}$ at $\sqrt{s}$ = 62,
130 and 200~GeV are 2.01, 2.25 and 2.43 ($\pm0.08$ error for each),
respectively, using the energy dependent parametrization
of the UA5~\cite{ua5} and CDF data
given in Ref.~\cite{abe}. For the $\sqrt{s}$ = 20~GeV data, which is outside
the energy range parametrized in~\cite{abe}, the average of ISR~\cite{isr-thome}
and FNAL~\cite{fnal} measurements summarized in Ref.~\cite{ua1} was assumed where
$n_{pp} = 1.4 \pm 0.12$.

Binary scattering parameter $x$ in Eq.~(\ref{Eq1}) was estimated by fitting
$(dN_{ch}/d\eta)/(0.5N_{part})$ data from STAR~\cite{starspec,molnar},
PHENIX~\cite{phenixdata}, and PHOBOS~\cite{phobos2002,phobos2006}
versus $\nu$ with Eq.~(\ref{Eq1}) rewritten as,
\bea
\frac{2}{N_{part}} \frac{d\bar{N}_{ch}}{d\eta} & = & n_{pp}[1+x(\nu-1)],
\label{Eq5}
\eea
assuming the above values for $n_{pp}$. The resulting values of $x$ are
0.07, 0.09, 0.09 and 0.13 ($\pm$0.03 errors for each) for the 20, 62, 130
and 200~GeV data, respectively.  Analysis of the 19.6~GeV Au-Au
data by the PHOBOS experiment~\cite{phobos2002} assumed $n_{pp} = 1.27 \pm 0.13$~\cite{isr-thome}
and obtained $x = 0.12$, leading to claims in the literature that hard
scattering contributions to multiplicity do not change with collision
energies from 20 to 200~GeV~\cite{phobos2002,phobos2006,loizides,stephans}.
A more recent compilation~\cite{ua1} of $n_{pp}$ measurements in this
energy range indicates a larger value of $n_{pp} = 1.4 \pm 0.12$.  Fitting
Eq.~(\ref{Eq5}) to the combined PHENIX~\cite{phenixdata} and
PHOBOS~\cite{phobos2002} 19.6~GeV data with $n_{pp} = 1.4$ resulted in a
smaller $x = 0.07$. PHOBOS 62~GeV Au-Au data~\cite{phobos2006}, when plotted
versus $\nu$, do not linearly extrapolate to $n_{pp}$ in contrast to 
what is expected from 
Fig.~4 of Ref.~\cite{phobos2006} and Fig.~1 of Ref.~\cite{stephans} from
the PHOBOS experiment.
However, 62~GeV Au-Au results from the STAR experiment linearly extrapolate to $n_{pp}$ at $\nu=1$,
resulting in the fitted value $x = 0.09$ used here.  $x$ parameters from the three RHIC experiments
generally agree for the 130 and 200~GeV data; the PHOBOS values of
$x = 0.09$~\cite{phobos2002} and 0.13~\cite{phobos2002} at 130 and 200~GeV,
respectively, were confirmed and used here.
It is possible that parameter $x$
is affected by other processes
in addition to hard and semi-hard partonic scattering in the initial collision
stage where the latter mechanisms would be expected to follow a $\log\sqrt{s}$ dependence.

The variance of the multiplicity distribution ${\cal P}(N_{ch},\bar{N}_{ch})$
in Eq.~(\ref{Eq2}) is given by $a\bar{N}_{ch}$ where values of parameter
$a<1$ or $>1$ represent multiplicity fluctuation suppression or excess,
respectively, relative to pure statistical fluctuations ($a=1$). In general
a non-vanishing integral~\cite{clt,delsign,ptscale} of two-particle correlations~\cite{axialci}
over the acceptance requires $a \neq 1$.  In principle, correlation measures
of the type reported in Ref.~\cite{axialci} could be used to determine parameter $a$
using the relationship between fluctuations and correlations developed
in Refs.~\cite{clt,ptscale} and discussed in Appendix~A. 
On the other hand, the Kharzeev and Nardi two-component
multiplicity model with distribution ${\cal P}(N_{ch},\bar{N}_{ch})$
constitutes a phenomenology for describing event-wise multiplicity frequency
distribution data. The phenomenological approach based on fits to multiplicity
distribution data, shown in the next section,  was used to estimate parameter
$a$ in the present analysis. The uncertainty in the width of ${\cal P}(N_{ch},\bar{N}_{ch})$ was estimated by fitting the data. The uncertainty in the
analytic form of the multiplicity fluctuation distribution was accounted for
by comparing Monte
Carlo Glauber results assuming the Gaussian ${\cal P}(N_{ch},\bar{N}_{ch})$ in Eq.~(\ref{Eq2})
(with $a=1$) with results assuming a negative binomial distribution
(NBD)~\cite{nbd} in place of Eq.~(\ref{Eq2})
as explained in Appendix~A.

\section{Fits to 130 GeV Au-Au Data}
\label{Sec:datafit}

\begin{figure*}[t]
\includegraphics[keepaspectratio,width=6.3in]{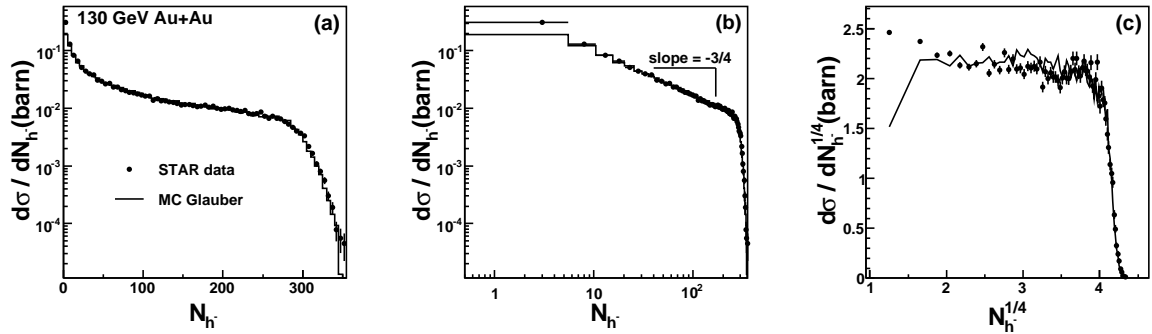}
\caption{\label{Fig1}
Negative hadron multiplicity frequency distributions
for Au-Au minimum-bias collisions at $\sqrt{s_{NN}}$ = 130~GeV normalized
to the total inelastic cross section in barns.
Left panel (a): semi-log plot comparing STAR data~\cite{starspec}
(solid dots) with the
Monte Carlo Glauber model fit (histogram). Middle panel (b): same
data and Monte Carlo fit on a log-log plot showing
the power-law dependence and the exponent (slope)
of approximately -3/4.
Right panel (c): same data (solid dots) and Monte Carlo fit (solid line) plotted as $d\sigma/dN_{h^-}^{1/4}$ versus
$N_{h^-}^{1/4}$.}
\end{figure*}

The 130~GeV Au-Au minimum-bias negatively charged hadron multiplicity frequency
distribution data from STAR~\cite{starspec} for 60K events
were fit by 
adjusting parameters $n_{pp}$ (for negative hadrons) and $a$ where
$x$ was fixed to 0.09 as discussed above. The acceptance was defined by transverse momentum
$(p_t) > 0.1$~GeV/$c$, $|\eta| < 0.5$, and $\Delta\phi = 2\pi$.
The fit (solid histogram) is shown in
the left panel of Fig.~\ref{Fig1}
in comparison with data (solid dots), where the optimum values of $n_{pp}$ and
$a$ are $1.110\pm0.004$ (consistent with $n_{pp} = 2.25\pm0.08$ for
charged particle yields) and $1.04\pm0.10$, respectively. Similar fits
(not shown) to the charged particle minimum-bias multiplicity distribution
from the same data set resulted in $a = 0.94\pm0.15$. Parameter $a$ was therefore
set to $1.0\pm0.2$ for all four energies.

In the TP analysis~\cite{powerlaw} it was shown that
minimum-bias multiplicity
frequency distributions for relativistic heavy ion collision
experiments and Monte Carlo Glauber models approximately follow
a power-law distribution.  This is illustrated by plotting
the data and MCG fit from the left panel of Fig.~\ref{Fig1} on log-log axes
in the middle panel.
Except near the end-points the data can be described to within 10\% with a
slope (exponent) of approximately $-3/4$.
The collision event yield is approximately proportional
to $N_{h^-}^{-3/4}$.
Therefore distribution $d\sigma/dN_{h^-}^{1/4}$
versus $N_{h^-}^{1/4}$ is approximately constant, where
\bea
\frac{d\sigma}{dN_{h^-}^{1/4}} & = & \frac{dN_{h^-}}{dN_{h^-}^{1/4}}
\frac{d\sigma}{dN_{h^-}} = 4N_{h^-}^{3/4}\frac{d\sigma}{dN_{h^-}}
\approx {\rm const.}
\eea
as shown in the right-hand panel of Fig.~\ref{Fig1} for
data (solid dots) and MCG fit (solid line). The power-law plotting format in the right-hand panel enables a more sensitive comparison between the model fit and data than in the usual semi-log format (left panel). The MCG model fit is consistent with the data except for the first two data points at low $N_{h^-}^{1/4}$. Except near the end-points the $d\sigma/dN_{h^-}^{1/4}$ data are
constant on $N_{h^-}^{1/4}$ to within an overall variation of 20\% in comparison to the
$d\sigma/dN_{h^-}$ data which span nearly two orders of magnitude within the
same multiplicity range.

The results in Fig.~\ref{Fig1} demonstrate the efficacy of the Monte Carlo Glauber model with
two-component multiplicity production for accurately describing the measured
multiplicity frequency distributions at RHIC.
Based on this outcome we conclude that the present model is reasonable to use in 
estimating centrality bin average quantities and their
systematic errors for RHIC data.

\section{Results and Error Analysis}
\label{Sec:results}

Ensembles of one-million, minimum-bias ({\em i.e.} random impact parameter)
Monte Carlo collisions were generated for each of the six systems studied here.
Centrality bin averaged quantities $\langle N_{ch} \rangle$,
$\langle b \rangle$, $\langle N_{part} \rangle$,
$\langle N_{bin} \rangle$ and $\nu$
are listed in Tables~\ref{TableII}-\ref{TableVII}
using the nominal parameter values
discussed above.
Statistical errors are typically 0.1-0.2\% and always $<0.5$\% of the
nominal bin averages and in all instances are much less than the systematic
errors discussed below.
Results in Table~\ref{TableII} for $\langle N_{part} \rangle$ for Au-Au collisions at 200~GeV
are systematically larger than published Monte Carlo 
Glauber predictions in Ref.~\cite{starmc} by approximately 5\% for peripheral
centrality bins. This systematic increase is primarily caused by multiplicity fluctuations in the present model
and basing centrality on $N_{ch}$ here rather than on $N_{part}$ as was done
in Ref.~\cite{starmc}. 

\begin{figure}[htb] 
\begin{center}
\includegraphics[keepaspectratio,width=3.0in]{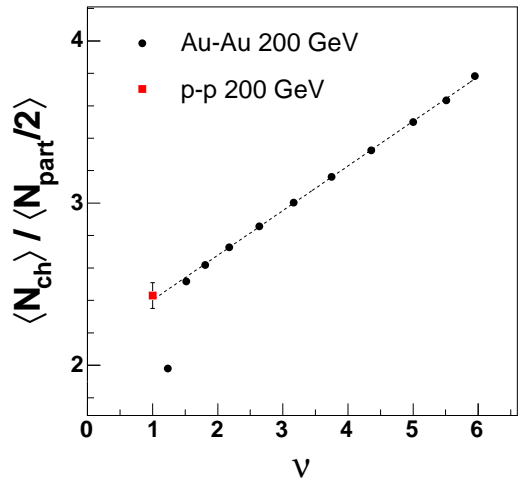}
\end{center}
\caption{\label{Fig2}
(color online) Monte Carlo Glauber results for
$\langle N_{ch} \rangle / \langle N_{part}/2 \rangle$ for
one-million Au-Au collisions at $\sqrt{s_{NN}}$ = 200~GeV
using the nominal parameters discussed in the text (solid dots).
Linear fit (dashed line) to the Au-Au results (excluding the 90-100\%
centrality bin) accurately extrapolates to the p-p multiplicity from
UA5~\cite{ua5,abe}. Monte Carlo data for Au-Au include statistical errors only which are
smaller than the symbols.}
\end{figure}

Average charged particle multiplicity per participant pair for $|\eta| < 0.5$
at mid-pseudorapidity as a function of centrality ($\nu$) is shown in
Fig.~\ref{Fig2} for Monte Carlo simulated Au-Au collisions at 
$\sqrt{s_{NN}}$ = 200~GeV (solid dots). The p-p limit, $n_{pp} = 2.43 \pm 0.08$,
is indicated by the solid square symbol.
The data display a linear dependence on $(\nu - 1)$ as expected from Eq.~(\ref{Eq5}) except for the most-peripheral 90-100\% centrality bin where multiplicity
fluctuations significantly reduce the average $N_{ch}$. The slope agrees with
$n_{pp} x = 0.32$; the linear extrapolation to the p-p limit is evident.

\subsection{Error Estimation Method}
\label{Subsec:errorestimate}

The sources of uncertainty which cause systematic error in the collision
geometry quantities can be organized into three categories.
The first includes fitting uncertainties that occur when the corrected,
multiplicity frequency distribution data are described by adjusting the
parameters of the multiplicity production model in Eqs.~(\ref{Eq1}-\ref{Eq2}). 
The second includes the uncertainties in the matter density, nucleon-nucleon
total inelastic cross section, and functional representation of the 
multiplicity fluctuation distribution ${\cal P}(N_{ch},\bar{N}_{ch})$. The
third includes the uncertainty in the corrected multiplicity frequency
distribution data $dN_{evt}/dN_{ch}$. The latter
arise from trigger inefficiency, collision
vertex finding inefficiency, background contamination, and particle
tajectory finding inefficiency.

Systematic errors in centrality bin average quantities due to fitting uncertainty were estimated by comparing the nominal results to that obtained
with parameters $x$ and $a$ in Eqs.~(\ref{Eq1}-\ref{Eq2}) individually varied
within their respective fitting errors (Secs.~\ref{Subsec:twocomponent} and \ref{Sec:datafit}). Uncertainty in parameter
$n_{pp}$ directly affects $\langle N_{ch} \rangle$ but does not affect $\langle b \rangle$, $\langle N_{part} \rangle$,
$\langle N_{bin} \rangle$ or $\nu$.

Error estimates due to uncertainties in the matter density radius and diffuseness parameters and the N-N inelastic cross section require a refitting of the $dN_{evt}/dN_{ch}$ distribution. This was accomplished via a phenomenological adjustment of the multiplicity production model described in Appendix~B. Fit recovery via $|\chi|^2$ minimization for MCG simulations with ample statistics (of order $10^6$ collisions) is computationally intensive. The method in Appendix~B is fast and accurate. Errors due to the uncertainty in the mathematical representation of ${\cal P}(N_{ch},\bar{N}_{ch})$ were estimated by comparing the nominal MCG results using the Gaussian distribution in Eq.~(\ref{Eq2}) with results assuming a negative binomial distribution~\cite{ua5,nbd} as explained in Appendix~A. 

Uncertainties in the $dN_{evt}/dN_{ch}$ data arising from trigger inefficiency, collision
vertex finding inefficiency, and background contamination mainly affect the
low multiplicity region. Minimum-bias trigger efficiencies at RHIC are
$92.2^{+2.5}_{-3.0}$\%~\cite{phenixdata,phenixspec},
$94\pm2$\%~\cite{starspec}, 96\%~\cite{brahmsspec}
and $97\pm3$\%~\cite{phobos2002,phobosspec,phobosspec3}. Trigger inefficiency
causes the lower $N_{ch}$ half-max point of the $dN_{evt}/dN_{ch}^{1/4}$
distribution to shift to larger $N_{ch}^{1/4}$. The position of this point for the
$dN_{evt}/dN_{ch}^{1/4}$ distribution when corrected for trigger inefficiency
(assuming the power-law behavior) has a relative uncertainty of about
$\pm$10\%.~\footnote{Trigger efficiencies are uncertain by about $\pm$2 to $\pm$3\%. The fractional
uncertainty in the lower half-max position is $\pm$(0.02 to 0.03)$(N_{ch,max}^{1/4} - N_{ch,min}^{1/4})/N_{ch,min}^{1/4} \approx 10$\% where
$N_{ch,min}^{1/4}$ and $N_{ch,max}^{1/4}$ are the lower and upper half-max points of the $dN_{evt}/dN_{ch}^{1/4}$ distribution.} However, knowledge that the lower half-max point is
constrained by p-p scattering allows the relative uncertainties of the
lower half-max point on $N_{ch}^{1/4}$ to be reduced to 1/4 of the uncertainties in $n_{pp}$,
or to $\pm$0.8\%, $\pm$0.9\%, $\pm$1\%, and $\pm$2\% for the 200, 130, 62
and 20~GeV data, respectively.

Primary collision vertex reconstruction efficiency is approximately 100\%
for events with $N_{ch}$ of order a few tens and greater but falls
precipitously for $N_{ch}<10$~\cite{calderon}. PHOBOS~\cite{phobos2002,phobosspec3}
and STAR~\cite{starspec,calderon} both report uncertainties in their overall
vertex finding efficiencies of about $\pm2$\% for minimum-bias
collisions. Vertex finding inefficiency increases the slope and half-max
position of the low multiplicity edge of the $dN_{evt}/dN_{ch}^{1/4}$ distribution.
An uncertainty of $\pm2$\% in the overall efficiency results in about
$\pm$10\% uncertainty in the low $N_{ch}^{1/4}$ half-max position.
However, the p-p data constrain this uncertainty. In this
analysis we assumed 100\% vertex finding efficiency for collisions producing
$N_{ch} \geq 14$~\cite{calderon}, or $N_{ch}^{1/4} \geq 1.93$, and allowed
the slope of the lower edge of the $dN_{evt}/dN_{ch}^{1/4}$ distribution to vary such that the half-max position varied by $\pm$0.8\%, $\pm$0.9\%, $\pm$1\%, and $\pm$2\% for the 200, 130, 62
and 20~GeV data, respectively, as in the preceding paragraph.

The principle sources of background contamination are from
ultra-peripheral two-photon interactions~\cite{upc}
and beam-gas collisions. The former process corresponds to coherent photon-photon
interactions which excite both nuclei, followed by neutron decay (which activates the minimum-bias trigger detectors) and accompanied by
resonance(s) production which decays into charged particles
transverse to the beam direction.  Transverse particle multiplicities from
UPC events are typically $\leq 2$ ({\em e.g.} $\rho$-meson
decay) and generally $\leq 4$~\cite{upc} for $|\eta|<1$ at midrapidity.
UPC backgrounds are therefore restricted to the most-peripheral
(90-100\%) centrality bin.  UPC yields should be approximately proportional to
$(Z_1 Z_2)^2$~\cite{upc} (charge numbers for colliding ions 1 and 2) whereas beam-gas
contamination should scale with beam current. Other background events,
{\em e.g.} mutual Coulomb dissociation processes such as $\gamma$ + A $\rightarrow ~ {\rm A}^{\star} 
~ \rightarrow$ B + n for both nuclei, can be eliminated by requiring minimum
transverse particle production. Remaining background events will appear near
the lower $N_{ch}$ edge of the $dN_{evt}/dN_{ch}^{1/4}$ distribution.

Estimates of background contamination in minimum-bias trigger data at RHIC
range from $1\pm1$\%~\cite{phenixspec} to 6\%~\cite{starmc,starspec}
overall for Au-Au at $\sqrt{s_{NN}}$ =130~GeV corresponding to 0-20\% and
60\%, respectively, of the hadronic collision event yield in the 90-100\% centrality
bin.  Most of the UPC events occur at $N_{ch} < n_{pp}$ and can be
eliminated by cuts on the number of transverse charged particles.
The remaining background contributions for $N_{ch} > n_{pp}$ are less than
the amounts listed above. For the present analysis background contamination
was assumed to be dominated by UPC events and to diminish in magnitude with
collision energy and $(Z_1 Z_2)^2$. In the present analysis background
contamination was applied to the nominal 90-100\% centrality bin assuming
3\%, 2\% and 1\% overall contamination levels in the Au-Au data at 
$\sqrt{s_{NN}}$ = 200, 130 and 62~GeV, respectively.
UPC contamination for Au-Au collisions at 20~GeV and for Cu-Cu at 200 and 62~GeV
was estimated to be negligible. 
However, calculations were done for the latter
three systems assuming a 1\% overall background contamination (10\%
contamination within the nominal 90-100\% centrality bin)
in order to provide a reference for further systematic error estimation.

The above sources of systematic uncertainty primarily affect the position, slope and shape of the $dN_{evt}/dN_{ch}^{1/4}$ distribution at the lower $N_{ch}^{1/4}$ end-point. These changes impact the MCG model which must describe those data and, in turn, the collision geometry quantities like $\langle N_{part} \rangle$. The phenomenological method in Appendix~B was used to estimate the changes in the collision geometry quantities relative to the nominal values. The latter differences were taken as the estimated errors.

Background levels in collider experiments vary significantly depending
on beam quality, beam current and interaction rate.
Excessive trigger backgrounds beyond
that considered here may result
from beam-gas interactions during periods of high integrated beam
currents. Collision event pile-up in the particle tracking detectors
during periods of high luminosity adversely affect collision vertex finding.  Either condition may be so severe as to
preclude access to the low
multiplicity range of the minimum-bias distribution. Even so, the power-law
dependence and p-p end-point constraints enable accurate centrality estimates
and collision geometry assignments to be made for the remaining minimum-bias data.

Charged particle trajectory reconstruction efficiencies in the large acceptance
tracking detectors at RHIC are typically
70 - 95\%~\cite{starspec,phenixdata,phobos2002} and
decrease approximately linearly with particle density in the detectors by
about 20\% from most-peripheral to most-central collisions~\cite{calderon}.
Uncertainties in the assumed track reconstruction efficiencies
were estimated by comparing corrected data for
$(dN_{ch}/d\eta|_{\eta=0})/(N_{part}/2)$ versus $N_{part}$
between the RHIC experiments for Au-Au collisions at 20~GeV~\cite{phenixdata,phobos2002}
130~GeV~\cite{starspec,phenixdata,phobos2002} and
200~GeV~\cite{phenixdata,phobos2002}.  The comparisons indicate
uncertainties in both the overall tracking efficiencies and in the dependence on particle density in the tracking detectors. The latter variation in efficiency from peripheral to central collisions is about $20\pm8$\%.

Overall changes in tracking efficiency are compensated in the MCG model by
multiplicative adjustments to parameter $n_{pp}$ and therefore have no effect on
the centrality measures reported here.
Changes in the assumed tracking efficiency dependence on $N_{ch}$ affect
the lower and upper end-point positions of the $dN_{evt}/dN_{ch}^{1/4}$
distribution and distort its shape. The distortions must be accounted for
by the MCG model in order to determine the net effect on the collision geometry measures. The effects of the 
8\% uncertainty in the multiplicity dependence of the trajectory reconstruction efficiency were estimated by
generalizing parameter $n_{pp}$ in Eq.~(\ref{Eq1}) to
$n_{pp}(1 + \alpha N_{part})$ where $\alpha = \pm 0.00023$ and
$\alpha N_{part} = \pm 0.08$ for most-central Au-Au collisions.  The same
value for $\alpha$ was assumed for Cu-Cu. Systematic errors were estimated
as the differences between the nominal centrality bin averages and those
obtained assuming $\alpha = \pm 0.00023$.  The choice
to maximize multiplicity variation
for most-central collisions is arbitrary. As a result systematic errors in centrality bin average
multiplicities were not included in Tables~\ref{TableII}-\ref{TableVII}.
However, this ambiguity does not affect the resulting systematic errors
in the other centrality measures reported here.

\subsection{Error Results}
\label{Subsec:errorresults}

The combined systematic errors (all components added in quadrature)
in both magnitude and relative percent (given in parentheses) are listed
in Tables~\ref{TableII} - \ref{TableVII} for the six collision systems
studied here. Impact parameter uncertainty is about $\pm(2\, {\rm to}\, 3)$\%
for Au-Au and $\pm(2\, {\rm to}\, 6)$\% for Cu-Cu.
Uncertainty in $\langle N_{part} \rangle$ is about $\pm(4\, {\rm to}\, 8)$\%
for Au-Au and $\pm(3\, {\rm to}\, 7)$\% for Cu-Cu for peripheral centralities,
reducing to about $\pm1$\% or less for central collisions where full
geometrical overlap of the two nuclei suppresses the dependence of
$N_{part}$ on variations in the nuclear surface geometry.  Systematic
errors in $\langle N_{bin} \rangle$ vary from about $\pm(4\, {\rm to}\, 12)$\%
for Au-Au collisions and $\pm(5\, {\rm to}\, 10)$\% for Cu-Cu
for central to peripheral collisions.
Errors in $\nu$
($\pm4$\% or less) are suppressed due to covariation of $\langle N_{part} \rangle$ and $\langle N_{bin} \rangle$.

Individual contributions to the systematic errors in collision geometry bin
averages $\langle b \rangle$, $\langle N_{part} \rangle$,
$\langle N_{bin} \rangle$ and
$\nu$ for Au-Au and Cu-Cu collisions at $\sqrt{s_{NN}}$ = 200~GeV are
summarized in Tables~\ref{TableVIII} and \ref{TableIX}, respectively.
Similar results were obtained for the other four collision systems.
Errors are given in percent where the
absolute values of positive and negative errors were averaged together.
The three numbers listed for each instance correspond
to the average errors within the 60-100\%, 20-60\%
and 0-20\% centrality bins, respectively.
The dominant errors are due to uncertainties in the matter density
and the N-N total inelastic cross section.
Errors due to uncertainties in the analytic form of the phenomenological
multiplicity fluctuation model are only significant for 
peripheral collisions. Errors due to uncertainties in the multiplicity dependent particle track reconstruction
efficiency and in the trigger and vertex reconstruction inefficiencies
are negligible when constrained by p-p data
as shown previously
by Trainor and Prindle~\cite{powerlaw}.

Percent uncertainties due to possible background contamination are
listed for all six collision systems in Table~\ref{TableX}.
The largest errors occur in the most-peripheral centrality bin as
expected.  The absolute magnitudes of the errors decline rapidly
with increasing centrality and are negligible for the centrality bins
not listed in the table.  Reference errors for Au-Au at 20~GeV and Cu-Cu
at $\sqrt{s_{NN}}$ = 200 and 62~GeV were based on an assumed 10\% background
contamination in the 90-100\% centrality bin.

Estimates of total systematic error when background contamination
differs from that assumed here can be obtained by removing the
error contributions in Table~\ref{TableX} from the
total errors in Tables~\ref{TableII} - \ref{TableIV} for Au-Au
collisions at 200, 130 and 62~GeV, respectively, and then adding (in
quadrature) the appropriately scaled background errors from Table~\ref{TableX}.
For Au-Au collisions at 20~GeV and Cu-Cu collisions at 200 and 62~GeV
the scaled errors from Table~\ref{TableX} should be combined in quadrature
with the total systematic errors in Tables~\ref{TableV} - \ref{TableVII}.
If much larger backgrounds
than those assumed here are encountered, then
the present Monte Carlo results should not be scaled; rather the errors
should be recalculated.

Overall, the systematic errors in collision geometry bin
averages for non-peripheral centralities are dominated by uncertainties in the nuclear geometry
and $\sigma_{inel}$. Errors in the more peripheral bins are
dominated by background contamination and ambiguities in the analytic form of the
multiplicity fluctuation model.

\section{Analytic Parametrizations}
\label{Sec:analytic}

The power-law description of heavy-ion collision centrality developed by Trainor and
Prindle~\cite{powerlaw} prescribes simple, analytic parametrizations of the $N_{part}$,
$N_{bin}$, $\nu$ and $N_{ch}$ dependences on total inelastic cross section
fraction $\sigma/\sigma_0$. For $N_{part}$ the running integral relation on
$(1 - \sigma/\sigma_0)$ is accurately given by~\cite{powerlaw}
\bea
(1 - \sigma/\sigma_0) & = & \frac{(N_{part}/2)^{\frac{1}{4}}
                            - (N_{part,min}/2)^{\frac{1}{4}}}
     {(N_{part,max}/2)^{\frac{1}{4}} - (N_{part,min}/2)^{\frac{1}{4}}},
\label{Eq7}
\eea
where
\bea
 (N_{part}/2)^{\frac{1}{4}}
 & = & 
  (N_{part,min}/2)^{\frac{1}{4}} \sigma/\sigma_0 
  +   (N_{part,max}/2)^{\frac{1}{4}} (1 - \sigma/\sigma_0).
\label{Eq8}
\eea
Similarly
\bea
 N_{bin}^{\frac{1}{6}} & = & N_{bin,min}^{\frac{1}{6}} \sigma/\sigma_0
+ N_{bin,max}^{\frac{1}{6}} (1 - \sigma/\sigma_0),
\label{Eq9}
\eea
where subscripts $min$ and $max$ refer respectively to the lower and upper
half-max end-point positions of the $d\sigma/d(N_{part}/2)^{1/4}$ and
$d\sigma/dN_{bin}^{1/6}$ distributions.

\begin{figure*}[t]
\begin{center}
\includegraphics[keepaspectratio,width=2.5in]{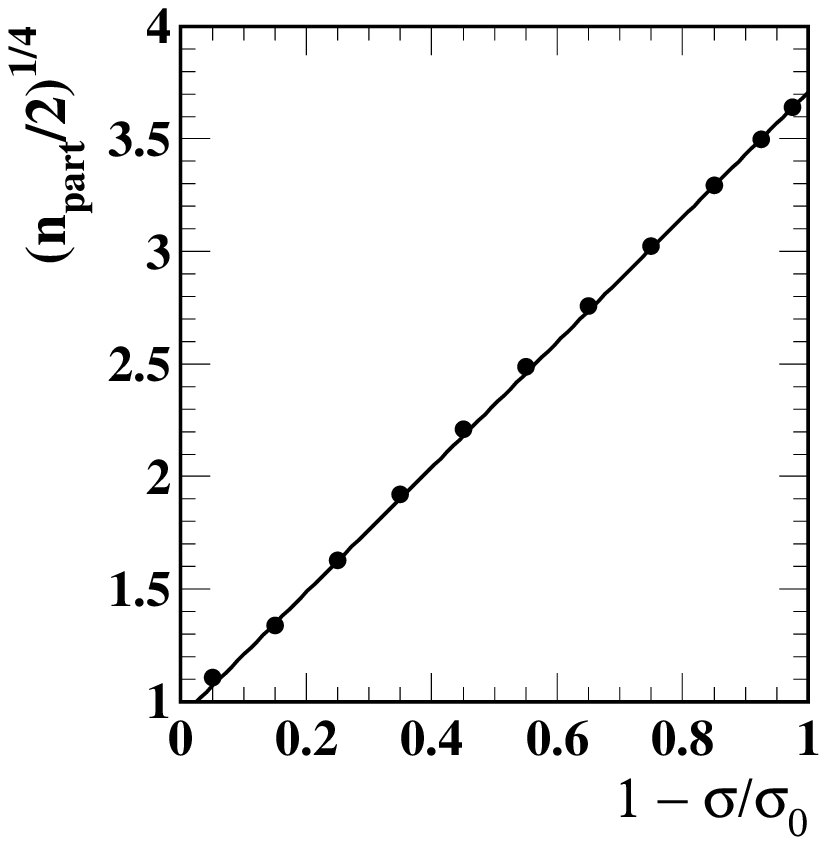}
\includegraphics[keepaspectratio,width=2.5in]{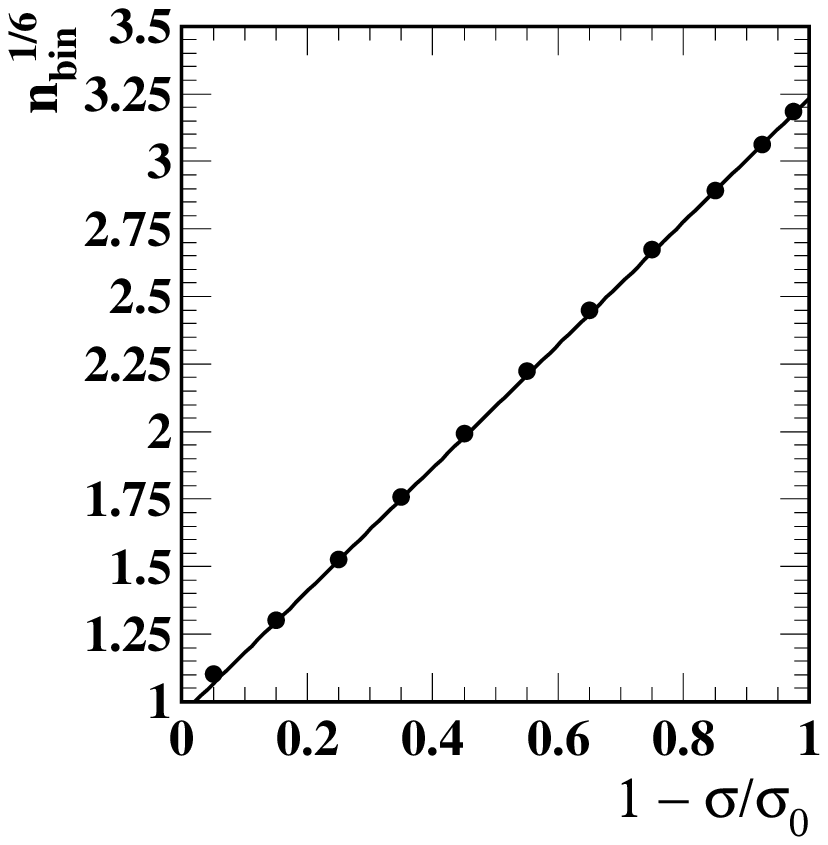}
\includegraphics[keepaspectratio,width=2.3in]{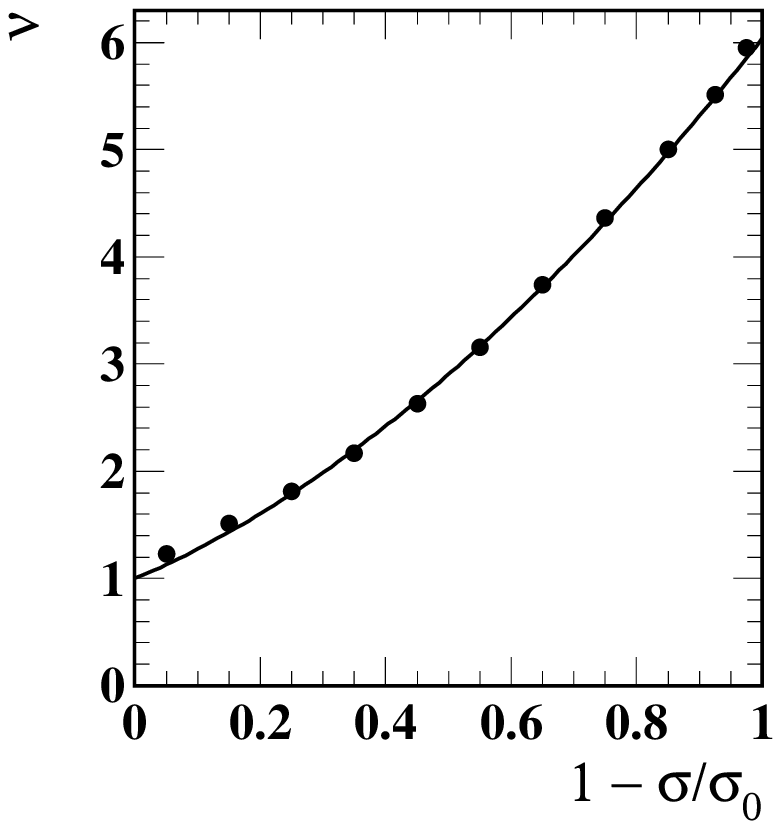}
\includegraphics[keepaspectratio,width=2.3in]{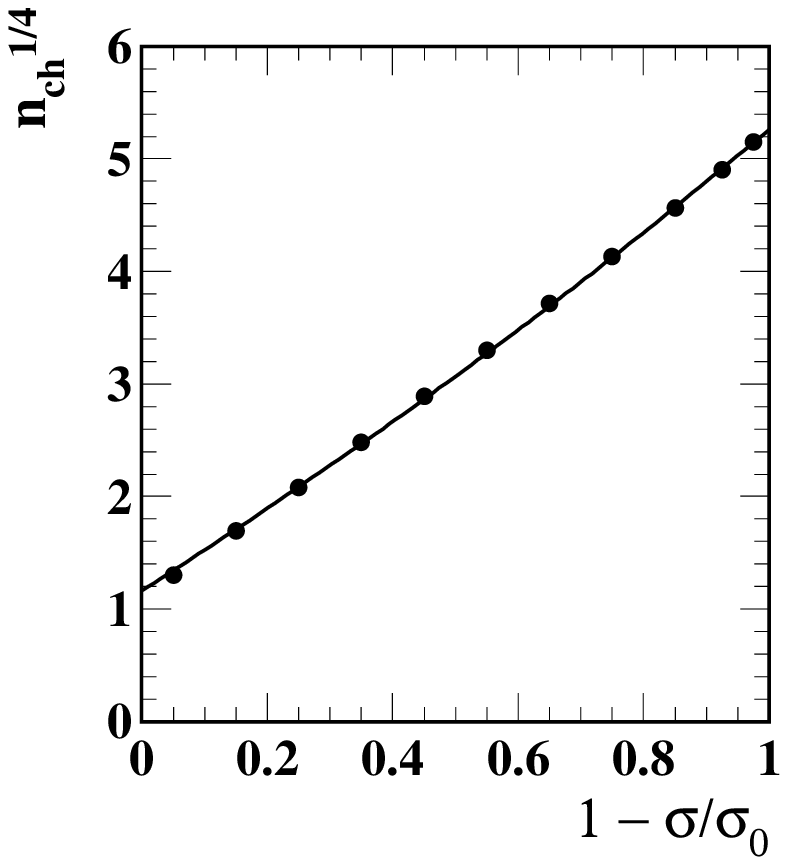}
\end{center}
\caption{\label{Fig3}
Comparison of Monte Carlo Glauber centrality quantities (solid dots) for Au-Au minimum-bias collisions at $\sqrt{s_{NN}}$ = 200~GeV from Table~\ref{TableII} with analytic parametrizations from Trainor and Prindle~\cite{powerlaw} based on the power-law description of heavy-ion collisions as discussed in the text. The four panels show from
upper-left to lower-right   
the dependences of $(N_{part}/2)^{1/4}$, $N_{bin}^{1/6}$, $\nu$ and
$N_{ch}^{1/4}$ on relative cross section fraction $(1 - \sigma/\sigma_0)$.}
\end{figure*}

The above parametrizations are shown as the solid lines in the two
upper      
panels of Fig.~\ref{Fig3} in comparison with the present MCG model results
(solid dots) from Table~\ref{TableII} for Au-Au collisions at $\sqrt{s_{NN}}$
= 200~GeV. The end-point parameters for the power-law parametrizations were $N_{part,min}/2 = 0.75$,
$N_{part,max}/2 = 189$, $N_{bin,min} = 0.75$ and $N_{bin,max} = 1144$.
The parametrization for $\nu = N_{bin}/(N_{part}/2)$ is compared with the MCG
result in the
lower-left     
panel. The average multiplicities using Eq.~(\ref{Eq5}),
the values for $n_{pp}$ and $x$ from Sec.~\ref{Subsec:twocomponent},
and the above parametrizations for $N_{part}$ and $\nu$ (solid curve) is
compared with the MCG result in the
lower-right    
panel. The simple power-law
parametrizations introduced in Ref.~\cite{powerlaw} quantitatively describe the
MCG results, thus confirming their utility and precision.

Calculation of the analytic power-law parametrizations require the upper half-max end-point values for $N_{part}$ and $N_{bin}$ from the MCG model. These quantities are listed in Table~\ref{TableXI} for the six collision systems studied here.

\section{Discussion}
\label{Sec:discussion}

Very peripheral collision data from the four RHIC experiments have generally
remained unpublished due to concerns with possibly large, and not well 
understood background contamination, trigger inefficiencies,
and primary collision
vertex finding inefficiencies.  Trainor and Prindle~\cite{powerlaw}
originally showed, and the present analysis confirms,
that the power-law dependence of the multiplicity frequency
distribution data together with knowledge of the proton-proton multiplicity
enables accurate centrality information to be obtained
in spite of these uncertainties.
Nevertheless, UPC events in principle cause the
lower end-point portion of the multiplicity frequency distribution
for minimum-bias A-A collisions to differ from the p-p limit
and these backgrounds will
contaminate the A-A data in the most-peripheral centrality bins. In this section we discuss
additional analysis methods using data available to the RHIC experiments
to minimize backgrounds and to
provide contamination level estimates for the accepted centrality bins.

Transverse particle production information via scintillators, silicon
detectors, and/or calorimetry are available in the RHIC experiments at the trigger level~\cite{startrig,phenixtrig,phobostrig,brahmstrig}.
These data record total energy deposition in the sensitive detector material for each (minimum bias) triggered
collision event including contributions
from A-A hadronic collisions, ultra-peripheral collisions, beam-gas and other
backgrounds, etc. but do not include particle track finding and primary
collision vertex finding inefficiencies. If the integrated yields from these
detectors, denoted as $SU\!M$, is approximately proportional to $N_{ch}$, then
the frequency distribution will follow an approximate power-law such that
the data can be usefully represented by $dN_{trig}/dSU\!M^{1/4}$ versus
$SU\!M^{1/4}$, where
$N_{trig}$ is the number of triggers.

\begin{figure}[htb]
\begin{center}
\includegraphics[keepaspectratio,width=4.0in]{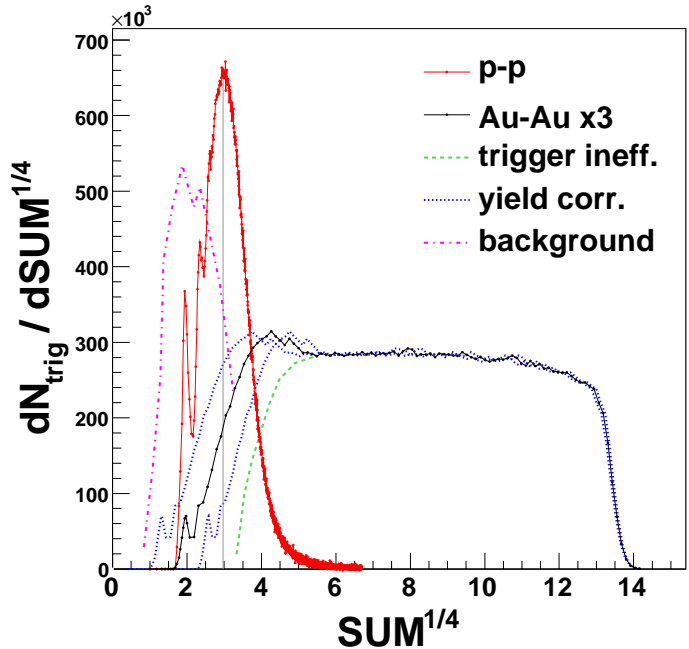}
\end{center}
\caption{\label{Fig4} (color online)
Monte Carlo simulation results for the minimum-bias trigger frequency distribution for
transverse particle production integrated trigger detector
yield quantity $SU\!M$ to the 1/4 power. Simulated yields are shown for 1M 200~GeV Au-Au (solid black curve, multiplied by 3)
and 1M proton-proton (solid red curve) collisions.
The yield for Au-Au collisions when
trigger inefficiency is large is illustrated by the lowest (right-most)
green dashed curve (hand-drawn sketch).
Over- and under-corrected yields are similarly illustrated by the upper (left-most)
and middle dotted blue lines, respectively.  Background contamination contributes
at lower multiplicity as illustrated by the dashed-dotted magenta curve (hand-drawn sketch).
}
\end{figure}

MCG simulations for the integrated yields from the STAR central trigger
barrel (CTB)~\cite{startrig} plastic scintillator detector ($|\eta| \leq 1$,
$2\pi$ azimuth coverage) were done for p-p and Au-Au minimum-bias collisions
at $\sqrt{s_{NN}}$ = 200~GeV. Details of this simulation are discussed in 
Appendix~C. The results for $dN_{trig}/dSU\!M^{1/4}$ versus $SU\!M^{1/4}$ 
are shown in Fig.~\ref{Fig4}. Both the p-p and Au-Au simulations reproduce
the general shapes of measured STAR CTB minimum-bias trigger yields~\cite{dunlop}. The
simulated Au-Au CTB yields follow the $SU\!M^{-3/4}$ power-law distribution
very well where the lower half-max point coincides with the mode of the p-p
distribution.  Trigger inefficiency in Au-Au causes a depletion at low
multiplicity as indicated by the green, dashed curve.  Over- or under-corrected
yields result in the blue dotted curves. From Fig.~\ref{Fig4} it is clear
how the p-p trigger yield can be used to constrain corrections for trigger inefficiency in A-A.

Background triggers appear at lower values of $SU\!M$ where UPC events
typically produce of order two charged particles at mid-rapidity compared
with the average from p-p collisions of about $5$ in $|\eta| \leq 1$ and
therefore appear as a peak or enhancement (dashed-dotted curve in
Fig.~\ref{Fig4}) below the mode of the p-p distribution. 
Comparisons between the measured p-p and Au-Au
$dN_{trig}/dSU\!M^{1/4}$ distributions and the power-law simulations for Au-Au
provide a reasonable means for defining additional event cuts to reduce
background contamination. This information can also be used to estimate the contamination level for
accepted events.  Measurements of the dependence of the
lower end-point region of the $dN_{trig}/dSU\!M^{1/4}$ 
distribution on integrated beam current, luminosity, and ion species 
would help disentangle background contributions from beam-gas collisions,
event pile-up, and UPC events. A similar power-law analysis
using reconstructed particle tracks, but without a primary collision vertex
requirement, would permit the vertex finding inefficiencies to be estimated
and corrected.

\section{Summary and Conclusions}
\label{Sec:summary}

A Monte Carlo Glauber and two-component multiplicity production model with fluctuations for high energy
heavy ion collisions
was used to describe minimum-bias multiplicity frequency distribution
data from RHIC.
Updated Woods-Saxon
radii and diffusivities for the point matter densities of $^{197}$Au and
$^{63}$Cu were determined using a combination of charge density measurements
from electron scattering and theoretical Hartree-Fock predictions.
The binary scattering parameter $x$ was estimated using a compilation
of available RHIC data from 20 to 200~GeV. The values for $x$ obtained here
display some energy dependence, although with large uncertainty.
The model was fitted
to the 130~GeV Au-Au minimum-bias data~\cite{starspec} in both the conventional
semi-log format and in the more sensitive power-law inspired format introduced
in Ref.~\cite{powerlaw}.
Quantitative agreement between the present model and data was obtained.

Systematic errors in centrality
bin averages for the geometrical quantities $\langle b \rangle$,
$\langle N_{part} \rangle$, $\langle N_{bin} \rangle$ and
$\nu = \langle N_{bin} \rangle/(\langle N_{part} \rangle /2)$ were
estimated based on
charged particle multiplicity for
$|\eta| \leq 0.5$ and full $2\pi$ azimuth acceptance.
The sources of systematic error considered here included
uncertainties associated with the model (nuclear density, nucleon-nucleon
inelastic total cross section,
multiplicity production model analytic form and parameters) and with the corrected
minimum-bias multiplicity frequency distribution data (background contamination,
trigger and collision vertex finding inefficiencies, and particle trajectory
reconstruction inefficiencies). 
The TP centrality analysis method
was applied to the MCG predictions and error analysis for
six collision systems relevant to the RHIC heavy ion program.

The analysis showed that significantly reduced errors in the 
collision geometry quantities result when the uncertainties in the
minimum-bias multiplicity frequency distribution data are constrained by the empirical power-law behavior and
the minimum-bias p-p collision data.  The reduction in errors is in agreement
with the original TP power-law analysis;
the resulting errors are in general smaller
than previous estimates~\cite{starmc,glauber-rev,starspec}
which did not utilize the power-law and p-p constraints.
The accuracy of simple parametrizations of centrality dependent quantities
developed by Trainor and Prindle~\cite{powerlaw} was confirmed.
We also discussed how particle production data at the trigger level
({\em e.g.} scintillator hits and calorimeter energy depositions)
can be used within the power-law
and p-p constraint methodology~\cite{powerlaw} to define additional event cuts which
minimize background contamination and which
enable the detrimental effects of trigger inefficiencies to be reduced. 

The systematic errors presented in this paper demonstrate the accuracy of collision geometry quantities which
can be achieved with RHIC data by exploiting the power-law behavior of the A-A
data and using constraints from minimum-bias p-p collision data. The results
presented here assume that
backgrounds from UPC events,
pileup, beam-gas collisions, etc. are not too large and that the minimum-bias
trigger and collision vertex reconstruction efficiencies are not too
small for peripheral collisions. It is intended that the MCG results
presented here will serve as a useful resource for the RHIC community
and that the analysis method developed by Trainor and Prindle~\cite{powerlaw}
and applied in this paper, together with the trigger data analysis method discussed here, 
will enable accurate description of, and better 
access to the heavy ion peripheral collision data from RHIC.

\ack
The authors gratefully acknowledge helpful discussions, detailed comments
and the plots in Fig.~\ref{Fig3} from Prof. T. A. Trainor (Univ. of Washington)
as well as helpful discussions and comments from Drs. J. C. Dunlop and
Zhangbu Xu (both from Brookhaven National Lab).
This research was supported in part by The U. S. Dept. of Energy Grant No.
DE-FG03-94ER40845.


\begin{table}[htb]
\begin{center}
\caption{\label{TableII}
Centrality-bin averaged collision geometry quantities and errors for Monte Carlo Glauber
Au-Au collisions at $\sqrt{s_{NN}}$ = 200~GeV using the nominal model
parameters discussed in the text.
Centrality was based on $N_{ch}$ for $|\eta|<0.5$ and full
$2\pi$ azimuthal acceptance.
Estimated positive and negative systematic errors are listed
as magnitudes and
percentages (in parentheses) of the mean values.
Errors in mean multiplicities were not
computed (see text).}
\vspace{0.1in}
\begin{tabular}{cccccc}  \br 
\multicolumn{6}{c}{{\bf Au-Au 200 GeV}}  \\
Centrality (\%) & $\langle N_{ch} \rangle$ & $\langle b \rangle$ &
$\langle N_{part} \rangle$ & $\langle N_{bin} \rangle$ & $\nu$ \\
\mr 
 \vspace{0.2cm}
 90-100 &   2.9 &$14.68^{+0.28(1.9)}_{-0.30(2.0)}$ &$  2.9^{+0.3(8.6)}_{-0.2(6.5
)}$ &$   1.8^{+ 0.2(12.4)}_{- 0.2(9.4)}$ &$1.23^{+0.04(3.0)}_{-0.03(2.4)}$\\
 \vspace{0.2cm}
 80-90 &   8.2 &$13.89^{+0.27(1.9)}_{-0.28(2.0)}$ &$  6.4^{+0.3(4.2)}_{-0.2(3.9)
}$ &$   4.9^{+ 0.3(6.5)}_{- 0.3(5.5)}$ &$1.51^{+0.03(2.3)}_{-0.03(1.9)}$\\
 \vspace{0.2cm}
 70-80 &  18.7 &$12.96^{+0.24(1.9)}_{-0.26(2.0)}$ &$ 14.1^{+0.6(4.2)}_{-0.5(3.3)
}$ &$  12.7^{+ 0.9(7.2)}_{- 0.7(5.6)}$ &$1.81^{+0.05(2.9)}_{-0.04(2.4)}$\\
 \vspace{0.2cm}
 60-70 &  37.8 &$12.03^{+0.23(1.9)}_{-0.24(2.0)}$ &$ 27.2^{+1.1(4.0)}_{-0.9(3.2)
}$ &$  29.5^{+ 2.3(7.9)}_{- 1.8(6.1)}$ &$2.17^{+0.08(3.8)}_{-0.07(3.0)}$\\
 \vspace{0.2cm}
 50-60 &  69.7 &$11.04^{+0.21(1.9)}_{-0.22(2.0)}$ &$ 47.5^{+1.4(2.9)}_{-1.1(2.4)
}$ &$  62.5^{+ 4.6(7.3)}_{- 3.8(6.2)}$ &$2.63^{+0.11(4.2)}_{-0.10(3.8)}$\\
 \vspace{0.2cm}
 40-50 & 118.5 &$ 9.98^{+0.18(1.8)}_{-0.20(2.0)}$ &$ 76.3^{+1.8(2.4)}_{-1.4(1.8)
}$ &$ 120.7^{+ 8.3(6.9)}_{- 6.7(5.5)}$ &$3.16^{+0.14(4.4)}_{-0.12(3.9)}$\\
 \vspace{0.2cm}
 30-40 & 190.2 &$ 8.80^{+0.18(2.0)}_{-0.16(1.8)}$ &$115.5^{+1.4(1.2)}_{-1.4(1.3)
}$ &$ 216.2^{+11.8(5.5)}_{-10.7(4.9)}$ &$3.74^{+0.16(4.2)}_{-0.14(3.8)}$\\
 \vspace{0.2cm}
 20-30 & 291.5 &$ 7.43^{+0.14(1.9)}_{-0.13(1.7)}$ &$167.1^{+1.2(0.7)}_{-1.8(1.1)
}$ &$ 364.1^{+16.7(4.6)}_{-17.2(4.7)}$ &$4.36^{+0.17(3.9)}_{-0.16(3.7)}$\\
 \vspace{0.2cm}
 10-20 & 433.5 &$ 5.75^{+0.11(1.9)}_{-0.11(1.9)}$ &$234.8^{+1.2(0.5)}_{-1.6(0.7)
}$ &$ 586.9^{+24.9(4.2)}_{-26.1(4.4)}$ &$5.00^{+0.19(3.8)}_{-0.19(3.8)}$\\
 \vspace{0.2cm}
  5-10 & 577.6 &$ 4.05^{+0.08(1.9)}_{-0.06(1.5)}$ &$299.7^{+0.8(0.3)}_{-1.5(0.5)
}$ &$ 826.0^{+32.3(3.9)}_{-34.7(4.2)}$ &$5.51^{+0.20(3.7)}_{-0.21(3.7)}$\\
 \vspace{0.2cm}
  0-5 & 705.6 &$ 2.30^{+0.06(2.8)}_{-0.05(2.0)}$ &$350.6^{+1.7(0.5)}_{-2.0(0.6)}
$ &$1043.6^{+43.2(4.1)}_{-44.1(4.2)}$ &$5.95^{+0.23(3.8)}_{-0.23(3.8)}$\\
\br 
\end{tabular}
\end{center}
\end{table}

\begin{table*}[t]
\begin{center}
\caption{\label{TableIII}
Same as Table~\ref{TableII} except for Au-Au collisions at
$\sqrt{s_{NN}}$ = 130~GeV.}
\vspace{0.1in}
\begin{tabular}{cccccc}  \hline \hline
\multicolumn{6}{c}{{\bf Au-Au 130 GeV}}  \\
Centrality (\%) & $\langle N_{ch} \rangle$ & $\langle b \rangle$ &
$\langle N_{part} \rangle$ & $\langle N_{bin} \rangle$ & $\nu$ \\
\hline
 \vspace{0.2cm}
 90-100 &   2.7 &$14.61^{+0.30(2.1)}_{-0.30(2.0)}$ &$  3.0^{+0.2(7.8)}_{-0.2(6.2
)}$ &$   1.8^{+ 0.2(11.1)}_{- 0.2(8.7)}$ &$1.24^{+0.04(2.9)}_{-0.03(2.3)}$\\
 \vspace{0.2cm}
 80-90 &   7.4 &$13.83^{+0.27(1.9)}_{-0.27(2.0)}$ &$  6.4^{+0.3(4.2)}_{-0.2(3.5)
}$ &$   4.8^{+ 0.3(6.4)}_{- 0.2(5.0)}$ &$1.50^{+0.03(2.1)}_{-0.03(1.7)}$\\
 \vspace{0.2cm}
 70-80 &  16.6 &$12.90^{+0.25(1.9)}_{-0.25(2.0)}$ &$ 14.0^{+0.6(4.2)}_{-0.5(3.6)
}$ &$  12.5^{+ 0.9(7.2)}_{- 0.7(5.9)}$ &$1.78^{+0.05(2.9)}_{-0.04(2.4)}$\\
 \vspace{0.2cm}
 60-70 &  33.2 &$11.97^{+0.24(2.0)}_{-0.24(2.0)}$ &$ 27.0^{+1.1(4.0)}_{-0.9(3.5)
}$ &$  28.7^{+ 2.3(7.9)}_{- 1.9(6.5)}$ &$2.13^{+0.08(3.8)}_{-0.07(3.1)}$\\
 \vspace{0.2cm}
 50-60 &  60.1 &$10.99^{+0.22(2.0)}_{-0.22(2.0)}$ &$ 47.0^{+1.5(3.2)}_{-1.3(2.8)
}$ &$  60.1^{+ 4.6(7.6)}_{- 3.8(6.3)}$ &$2.56^{+0.11(4.2)}_{-0.09(3.5)}$\\
 \vspace{0.2cm}
 40-50 & 100.4 &$ 9.93^{+0.19(1.9)}_{-0.19(1.9)}$ &$ 75.5^{+1.7(2.2)}_{-1.5(2.0)
}$ &$ 115.1^{+ 7.3(6.4)}_{- 6.8(5.9)}$ &$3.05^{+0.13(4.1)}_{-0.12(4.0)}$\\
 \vspace{0.2cm}
 30-40 & 157.8 &$ 8.76^{+0.15(1.8)}_{-0.18(2.1)}$ &$114.0^{+1.8(1.6)}_{-1.2(1.0)
}$ &$ 203.9^{+12.1(5.9)}_{- 9.3(4.6)}$ &$3.58^{+0.15(4.2)}_{-0.13(3.7)}$\\
 \vspace{0.2cm}
 20-30 & 237.7 &$ 7.41^{+0.11(1.4)}_{-0.16(2.2)}$ &$164.9^{+2.3(1.4)}_{-0.8(0.5)
}$ &$ 341.0^{+18.7(5.5)}_{-14.0(4.1)}$ &$4.14^{+0.17(4.1)}_{-0.15(3.7)}$\\
 \vspace{0.2cm}
 10-20 & 348.7 &$ 5.72^{+0.10(1.8)}_{-0.12(2.0)}$ &$232.4^{+1.6(0.7)}_{-1.0(0.4)
}$ &$ 548.1^{+25.1(4.6)}_{-22.2(4.0)}$ &$4.72^{+0.18(3.9)}_{-0.17(3.6)}$\\
 \vspace{0.2cm}
  5-10 & 459.5 &$ 4.05^{+0.05(1.1)}_{-0.08(2.0)}$ &$296.8^{+1.2(0.4)}_{-0.6(0.2)
}$ &$ 767.9^{+32.6(4.2)}_{-29.6(3.9)}$ &$5.17^{+0.20(3.9)}_{-0.19(3.7)}$\\
 \vspace{0.2cm}
  0-5 & 557.6 &$ 2.29^{+0.06(2.7)}_{-0.06(2.6)}$ &$348.2^{+2.0(0.6)}_{-1.9(0.5)}
$ &$ 968.0^{+42.3(4.4)}_{-40.4(4.2)}$ &$5.56^{+0.22(3.9)}_{-0.21(3.8)}$\\
\hline \hline
\end{tabular}
\end{center}
\end{table*}

\begin{table*}[t]
\begin{center}
\caption{\label{TableIV}
Same as Table~\ref{TableII} except for Au-Au collisions at
$\sqrt{s_{NN}}$ = 62~GeV.}
\vspace{0.1in}
\begin{tabular}{cccccc}  \hline \hline
\multicolumn{6}{c}{{\bf Au-Au 62 GeV}}  \\
Centrality (\%) & $\langle N_{ch} \rangle$ & $\langle b \rangle$ &
$\langle N_{part} \rangle$ & $\langle N_{bin} \rangle$ & $\nu$ \\
\hline
 \vspace{0.2cm}
 90-100 &   2.4 &$14.52^{+0.29(2.0)}_{-0.29(2.0)}$ &$  3.1^{+0.1(4.4)}_{-0.1(4.1
)}$ &$   1.9^{+ 0.1(6.3)}_{- 0.1(5.5)}$ &$1.24^{+0.02(1.9)}_{-0.02(1.4)}$\\
 \vspace{0.2cm}
 80-90 &   6.6 &$13.77^{+0.25(1.8)}_{-0.29(2.1)}$ &$  6.4^{+0.3(4.9)}_{-0.2(2.8)
}$ &$   4.8^{+ 0.3(7.1)}_{- 0.2(4.3)}$ &$1.49^{+0.03(2.3)}_{-0.02(1.5)}$\\
 \vspace{0.2cm}
 70-80 &  14.6 &$12.83^{+0.24(1.8)}_{-0.25(1.9)}$ &$ 13.9^{+0.6(4.4)}_{-0.4(3.1)
}$ &$  12.1^{+ 0.9(7.6)}_{- 0.6(5.2)}$ &$1.75^{+0.05(3.0)}_{-0.04(2.2)}$\\
 \vspace{0.2cm}
 60-70 &  29.0 &$11.91^{+0.21(1.7)}_{-0.24(2.0)}$ &$ 26.6^{+1.2(4.4)}_{-0.7(2.7)
}$ &$  27.5^{+ 2.2(8.1)}_{- 1.5(5.5)}$ &$2.07^{+0.07(3.6)}_{-0.06(3.0)}$\\
 \vspace{0.2cm}
 50-60 &  52.3 &$10.94^{+0.19(1.7)}_{-0.22(2.0)}$ &$ 46.2^{+1.5(3.3)}_{-1.1(2.3)
}$ &$  56.8^{+ 4.2(7.4)}_{- 3.2(5.6)}$ &$2.46^{+0.10(4.1)}_{-0.09(3.5)}$\\
 \vspace{0.2cm}
 40-50 &  87.1 &$ 9.88^{+0.17(1.7)}_{-0.18(1.8)}$ &$ 74.2^{+1.9(2.6)}_{-1.1(1.5)
}$ &$ 107.8^{+ 7.4(6.9)}_{- 5.3(4.9)}$ &$2.90^{+0.12(4.3)}_{-0.10(3.6)}$\\
 \vspace{0.2cm}
 30-40 & 136.9 &$ 8.71^{+0.15(1.7)}_{-0.17(1.9)}$ &$112.4^{+2.0(1.8)}_{-1.3(1.2)
}$ &$ 190.2^{+11.5(6.0)}_{- 8.7(4.6)}$ &$3.38^{+0.14(4.2)}_{-0.12(3.7)}$\\
 \vspace{0.2cm}
 20-30 & 205.5 &$ 7.36^{+0.12(1.6)}_{-0.15(2.0)}$ &$162.6^{+2.0(1.2)}_{-0.9(0.6)
}$ &$ 315.2^{+16.6(5.3)}_{-12.3(3.9)}$ &$3.88^{+0.16(4.1)}_{-0.14(3.5)}$\\
 \vspace{0.2cm}
 10-20 & 300.7 &$ 5.69^{+0.09(1.6)}_{-0.12(2.1)}$ &$229.2^{+1.9(0.8)}_{-1.0(0.4)
}$ &$ 503.6^{+23.6(4.7)}_{-20.4(4.0)}$ &$4.39^{+0.17(3.9)}_{-0.17(3.8)}$\\
 \vspace{0.2cm}
  5-10 & 395.5 &$ 4.02^{+0.06(1.4)}_{-0.07(1.8)}$ &$293.3^{+1.3(0.5)}_{-0.7(0.2)
}$ &$ 703.5^{+30.6(4.3)}_{-28.3(4.0)}$ &$4.80^{+0.19(3.9)}_{-0.18(3.8)}$\\
 \vspace{0.2cm}
  0-5 & 480.8 &$ 2.28^{+0.05(2.4)}_{-0.06(2.4)}$ &$345.0^{+1.8(0.5)}_{-2.0(0.6)}
$ &$ 884.9^{+38.8(4.4)}_{-39.5(4.5)}$ &$5.13^{+0.20(4.0)}_{-0.21(4.0)}$\\
\hline \hline
\end{tabular}
\end{center}
\end{table*}

\begin{table*}[t]
\begin{center}
\caption{\label{TableV}
Same as Table~\ref{TableII} except for Au-Au collisions at
$\sqrt{s_{NN}}$ = 20~GeV.}
\vspace{0.1in}
\begin{tabular}{cccccc}  \hline \hline
\multicolumn{6}{c}{{\bf Au-Au 20 GeV}}  \\
Centrality (\%) & $\langle N_{ch} \rangle$ & $\langle b \rangle$ &
$\langle N_{part} \rangle$ & $\langle N_{bin} \rangle$ & $\nu$ \\
\hline
 \vspace{0.2cm}
 90-100 &   1.9 &$14.38^{+0.27(1.9)}_{-0.30(2.1)}$ &$  3.4^{+0.1(3.9)}_{-0.1(1.9
)}$ &$   2.2^{+ 0.1(6.0)}_{- 0.1(3.3)}$ &$1.29^{+0.02(1.9)}_{-0.02(1.3)}$\\
 \vspace{0.2cm}
 80-90 &   4.8 &$13.65^{+0.27(2.0)}_{-0.27(2.0)}$ &$  6.9^{+0.4(6.3)}_{-0.4(5.3)
}$ &$   5.2^{+ 0.4(8.5)}_{- 0.3(6.7)}$ &$1.51^{+0.04(2.4)}_{-0.03(1.9)}$\\
 \vspace{0.2cm}
 70-80 &  10.5 &$12.72^{+0.24(1.9)}_{-0.27(2.1)}$ &$ 14.6^{+0.9(6.0)}_{-0.6(4.2)
}$ &$  12.8^{+ 1.2(9.1)}_{- 0.7(5.8)}$ &$1.75^{+0.06(3.3)}_{-0.04(2.1)}$\\
 \vspace{0.2cm}
 60-70 &  20.5 &$11.79^{+0.22(1.9)}_{-0.26(2.2)}$ &$ 27.6^{+1.5(5.5)}_{-1.1(4.1)
}$ &$  28.4^{+ 2.6(9.3)}_{- 1.9(6.7)}$ &$2.06^{+0.08(4.0)}_{-0.06(3.1)}$\\
 \vspace{0.2cm}
 50-60 &  36.2 &$10.82^{+0.20(1.8)}_{-0.23(2.1)}$ &$ 47.3^{+2.2(4.5)}_{-1.5(3.2)
}$ &$  57.4^{+ 4.9(8.5)}_{- 3.5(6.1)}$ &$2.42^{+0.10(4.2)}_{-0.09(3.5)}$\\
 \vspace{0.2cm}
 40-50 &  59.2 &$ 9.78^{+0.19(1.9)}_{-0.20(2.1)}$ &$ 75.2^{+2.6(3.5)}_{-1.9(2.5)
}$ &$ 106.6^{+ 8.2(7.6)}_{- 6.0(5.6)}$ &$2.83^{+0.12(4.4)}_{-0.11(3.7)}$\\
 \vspace{0.2cm}
 30-40 &  91.6 &$ 8.62^{+0.16(1.9)}_{-0.18(2.1)}$ &$113.2^{+3.1(2.7)}_{-2.2(1.9)
}$ &$ 185.1^{+12.3(6.6)}_{- 9.6(5.2)}$ &$3.27^{+0.14(4.2)}_{-0.13(3.9)}$\\
 \vspace{0.2cm}
 20-30 & 135.9 &$ 7.27^{+0.15(2.1)}_{-0.15(2.1)}$ &$163.3^{+3.0(1.8)}_{-2.5(1.5)
}$ &$ 304.2^{+16.7(5.5)}_{-14.9(4.9)}$ &$3.72^{+0.15(4.0)}_{-0.14(3.8)}$\\
 \vspace{0.2cm}
 10-20 & 196.1 &$ 5.61^{+0.11(2.0)}_{-0.12(2.2)}$ &$229.1^{+2.8(1.2)}_{-2.3(1.0)
}$ &$ 479.5^{+24.6(5.1)}_{-22.1(4.6)}$ &$4.19^{+0.17(4.1)}_{-0.16(3.9)}$\\
 \vspace{0.2cm}
  5-10 & 255.4 &$ 3.96^{+0.07(1.8)}_{-0.08(2.1)}$ &$292.2^{+2.2(0.8)}_{-2.1(0.7)
}$ &$ 664.2^{+31.6(4.8)}_{-30.9(4.7)}$ &$4.55^{+0.19(4.1)}_{-0.18(4.0)}$\\
 \vspace{0.2cm}
  0-5 & 309.2 &$ 2.29^{+0.07(3.2)}_{-0.07(3.1)}$ &$342.2^{+2.5(0.7)}_{-2.7(0.8)}
$ &$ 825.9^{+38.2(4.6)}_{-38.5(4.7)}$ &$4.83^{+0.19(4.0)}_{-0.20(4.1)}$\\
\hline \hline
\end{tabular}
\end{center}
\end{table*}

\begin{table*}[t]
\begin{center}
\caption{\label{TableVI}
Same as Table~\ref{TableII} except for Cu-Cu collisions at
$\sqrt{s_{NN}}$ = 200~GeV.}
\vspace{0.1in}
\begin{tabular}{cccccc}  \hline \hline
\multicolumn{6}{c}{{\bf Cu-Cu 200 GeV}}  \\
Centrality (\%) & $\langle N_{ch} \rangle$ & $\langle b \rangle$ &
$\langle N_{part} \rangle$ & $\langle N_{bin} \rangle$ & $\nu$ \\
\hline
 \vspace{0.2cm}
 90-100 &   2.2 &$10.18^{+0.22(2.1)}_{-0.22(2.1)}$ &$  2.6^{+0.2(7.0)}_{-0.2(7.1
)}$ &$   1.5^{+ 0.2(10.8)}_{- 0.2(10.6)}$ &$1.18^{+0.04(3.1)}_{-0.04(3.0)}$\\
 \vspace{0.2cm}
 80-90 &   4.7 &$ 9.79^{+0.22(2.2)}_{-0.20(2.0)}$ &$  3.8^{+0.2(4.5)}_{-0.2(4.7)
}$ &$   2.6^{+ 0.2(6.6)}_{- 0.2(6.7)}$ &$1.35^{+0.02(1.8)}_{-0.02(1.8)}$\\
 \vspace{0.2cm}
 70-80 &   8.4 &$ 9.10^{+0.20(2.2)}_{-0.19(2.1)}$ &$  6.6^{+0.2(2.3)}_{-0.2(2.8)
}$ &$   5.0^{+ 0.2(4.1)}_{- 0.2(4.4)}$ &$1.53^{+0.03(1.7)}_{-0.03(1.8)}$\\
 \vspace{0.2cm}
 60-70 &  14.3 &$ 8.40^{+0.19(2.2)}_{-0.18(2.1)}$ &$ 10.9^{+0.3(2.6)}_{-0.3(2.9)
}$ &$   9.3^{+ 0.4(4.5)}_{- 0.5(4.9)}$ &$1.71^{+0.03(2.0)}_{-0.04(2.2)}$\\
 \vspace{0.2cm}
 50-60 &  23.3 &$ 7.68^{+0.17(2.2)}_{-0.14(1.8)}$ &$ 17.2^{+0.4(2.1)}_{-0.5(2.6)
}$ &$  16.7^{+ 0.7(4.3)}_{- 0.8(5.1)}$ &$1.94^{+0.05(2.4)}_{-0.05(2.5)}$\\
 \vspace{0.2cm}
 40-50 &  36.6 &$ 6.92^{+0.14(2.0)}_{-0.14(2.0)}$ &$ 26.1^{+0.5(1.8)}_{-0.6(2.4)
}$ &$  28.9^{+ 1.3(4.5)}_{- 1.5(5.2)}$ &$2.21^{+0.06(2.8)}_{-0.06(2.8)}$\\
 \vspace{0.2cm}
 30-40 &  55.4 &$ 6.10^{+0.13(2.1)}_{-0.13(2.2)}$ &$ 38.1^{+0.6(1.7)}_{-0.8(2.0)
}$ &$  48.2^{+ 2.3(4.8)}_{- 2.5(5.1)}$ &$2.53^{+0.08(3.1)}_{-0.08(3.1)}$\\
 \vspace{0.2cm}
 20-30 &  81.5 &$ 5.15^{+0.10(2.0)}_{-0.11(2.2)}$ &$ 53.9^{+0.8(1.5)}_{-0.9(1.7)
}$ &$  77.9^{+ 3.9(5.0)}_{- 3.9(5.0)}$ &$2.89^{+0.10(3.4)}_{-0.10(3.4)}$\\
 \vspace{0.2cm}
 10-20 & 117.4 &$ 3.97^{+0.09(2.2)}_{-0.07(1.9)}$ &$ 74.4^{+0.8(1.1)}_{-1.1(1.5)
}$ &$ 123.0^{+ 5.6(4.6)}_{- 6.1(4.9)}$ &$3.31^{+0.11(3.4)}_{-0.12(3.5)}$\\
 \vspace{0.2cm}
  5-10 & 152.2 &$ 2.80^{+0.05(1.9)}_{-0.05(1.9)}$ &$ 93.2^{+0.8(0.8)}_{-1.0(1.1)
}$ &$ 169.9^{+ 7.4(4.3)}_{- 7.9(4.7)}$ &$3.65^{+0.13(3.6)}_{-0.13(3.6)}$\\
 \vspace{0.2cm}
  0-5 & 184.8 &$ 1.75^{+0.11(6.5)}_{-0.10(5.9)}$ &$106.5^{+1.5(1.4)}_{-1.6(1.5)}
$ &$ 211.4^{+10.1(4.8)}_{-10.5(5.0)}$ &$3.97^{+0.15(3.8)}_{-0.15(3.8)}$\\
\hline \hline
\end{tabular}
\end{center}
\end{table*}

\begin{table*}[t]
\begin{center}
\caption{\label{TableVII}
Same as Table~\ref{TableII} except for Cu-Cu collisions at
$\sqrt{s_{NN}}$ = 62~GeV.}
\vspace{0.1in}
\begin{tabular}{cccccc}  \hline \hline
\multicolumn{6}{c}{{\bf Cu-Cu 62 GeV}}  \\
Centrality (\%) & $\langle N_{ch} \rangle$ & $\langle b \rangle$ &
$\langle N_{part} \rangle$ & $\langle N_{bin} \rangle$ & $\nu$ \\
\hline
 \vspace{0.2cm}
 90-100 &   1.8 &$10.00^{+0.21(2.1)}_{-0.21(2.1)}$ &$  2.8^{+0.1(3.4)}_{-0.1(2.7
)}$ &$   1.7^{+ 0.1(5.1)}_{- 0.1(4.4)}$ &$1.20^{+0.02(1.4)}_{-0.02(1.3)}$\\
 \vspace{0.2cm}
 80-90 &   3.8 &$ 9.65^{+0.20(2.1)}_{-0.20(2.1)}$ &$  3.8^{+0.1(2.9)}_{-0.1(3.3)
}$ &$   2.6^{+ 0.1(4.2)}_{- 0.1(4.1)}$ &$1.34^{+0.02(1.1)}_{-0.02(1.3)}$\\
 \vspace{0.2cm}
 70-80 &   6.7 &$ 8.97^{+0.21(2.3)}_{-0.17(1.9)}$ &$  6.5^{+0.1(2.2)}_{-0.2(3.2)
}$ &$   4.9^{+ 0.2(3.3)}_{- 0.2(4.7)}$ &$1.50^{+0.02(1.3)}_{-0.02(1.6)}$\\
 \vspace{0.2cm}
 60-70 &  11.2 &$ 8.27^{+0.17(2.1)}_{-0.16(2.0)}$ &$ 10.6^{+0.3(2.7)}_{-0.3(2.8)
}$ &$   8.8^{+ 0.4(4.6)}_{- 0.4(4.5)}$ &$1.66^{+0.03(2.0)}_{-0.03(1.8)}$\\
 \vspace{0.2cm}
 50-60 &  17.9 &$ 7.57^{+0.15(1.9)}_{-0.15(2.0)}$ &$ 16.7^{+0.4(2.2)}_{-0.5(2.9)
}$ &$  15.5^{+ 0.7(4.5)}_{- 0.8(5.2)}$ &$1.85^{+0.04(2.3)}_{-0.04(2.4)}$\\
 \vspace{0.2cm}
 40-50 &  27.7 &$ 6.82^{+0.14(2.0)}_{-0.13(1.9)}$ &$ 25.3^{+0.4(1.8)}_{-0.7(2.6)
}$ &$  26.3^{+ 1.1(4.3)}_{- 1.4(5.2)}$ &$2.08^{+0.05(2.6)}_{-0.06(2.7)}$\\
 \vspace{0.2cm}
 30-40 &  41.3 &$ 6.00^{+0.13(2.1)}_{-0.11(1.8)}$ &$ 36.8^{+0.5(1.5)}_{-0.8(2.2)
}$ &$  43.2^{+ 1.9(4.3)}_{- 2.2(5.1)}$ &$2.35^{+0.07(3.0)}_{-0.07(3.1)}$\\
 \vspace{0.2cm}
 20-30 &  60.0 &$ 5.06^{+0.11(2.1)}_{-0.08(1.6)}$ &$ 52.1^{+0.5(1.0)}_{-1.0(1.9)
}$ &$  69.0^{+ 2.5(3.6)}_{- 3.5(5.1)}$ &$2.65^{+0.07(2.7)}_{-0.09(3.3)}$\\
 \vspace{0.2cm}
 10-20 &  84.9 &$ 3.91^{+0.07(1.8)}_{-0.06(1.6)}$ &$ 71.8^{+0.7(1.0)}_{-1.0(1.4)
}$ &$ 106.8^{+ 4.5(4.2)}_{- 5.3(5.0)}$ &$2.97^{+0.09(3.1)}_{-0.11(3.6)}$\\
 \vspace{0.2cm}
  5-10 & 108.9 &$ 2.76^{+0.05(1.7)}_{-0.03(1.3)}$ &$ 90.1^{+0.8(0.9)}_{-1.1(1.2)
}$ &$ 145.7^{+ 6.5(4.5)}_{- 6.9(4.7)}$ &$3.23^{+0.12(3.6)}_{-0.12(3.6)}$\\
 \vspace{0.2cm}
  0-5 & 132.2 &$ 1.77^{+0.10(5.6)}_{-0.08(4.7)}$ &$103.2^{+1.4(1.3)}_{-1.5(1.5)}
$ &$ 178.4^{+ 9.0(5.0)}_{- 9.1(5.1)}$ &$3.46^{+0.14(4.0)}_{-0.14(3.9)}$\\
\hline \hline
\end{tabular}
\end{center}
\end{table*}

\begin{table}[htb]
\caption{\label{TableVIII}
Absolute values of individual error contributions (in percent) to collision geometry bin average
quantities for Au-Au collisions at $\sqrt{s_{NN}}$ = 200~GeV
relative to the reference values in Table~\ref{TableII} as explained in the text. Errors are
denoted as $\Delta \langle b \rangle$, etc. The left-most column lists the
sources of uncertainty. Average errors for centrality bins 60-100\%, 20-60\%
and 0-20\% are listed from left to right, respectively, for each instance. }
\begin{indented}
\vspace{0.1in}
\item[]\begin{tabular}{@{}ccccc} \br 
Error & \multicolumn{4}{c}{{\bf Au-Au 200 GeV Errors}} \\
Source & $\Delta \langle b \rangle$(\%) & $\Delta \langle N_{part} \rangle$(\%) & $\Delta \langle N_{bin} \rangle$(\%) & $\Delta \nu$(\%) \\
\mr 
$c_{pt,m}$ & 0.9, 1.0, 0.8 \hspace{0.1in} & 0.1, 0.1, 0.2 \hspace{0.1in} &
             0.1, 0.9, 2.2 \hspace{0.1in} & 0.0, 0.9, 2.0 \\
$z_{pt,m}$ & 1.7, 1.6, 1.5 \hspace{0.1in} & 3.2, 1.6, 0.3 \hspace{0.1in} &
             5.2, 5.1, 2.6 \hspace{0.1in} & 2.1, 3.5, 2.3 \\
$\sigma_{inel}$ & 0.1, 0.1, 0.0 \hspace{0.1in} & 0.4, 0.5, 0.3 \hspace{0.1in} &
                  1.0, 2.0, 2.5 \hspace{0.1in} & 0.6, 1.5, 2.2 \\
$x$        & 0.0, 0.0, 0.1 \hspace{0.1in} & 0.2, 0.0, 0.0 \hspace{0.1in} &
             0.3, 0.0, 0.0 \hspace{0.1in} & 0.2, 0.1, 0.0 \\
$a$        & 0.0, 0.0, 0.2 \hspace{0.1in} & 0.3, 0.1, 0.0 \hspace{0.1in} &
             0.4, 0.2, 0.1 \hspace{0.1in} & 0.1, 0.1, 0.0 \\
$\cal P$-form & 0.1, 0.0, 0.7 \hspace{0.1in} & 2.2, 0.1, 0.2 \hspace{0.1in} &
             3.5, 0.3, 0.3 \hspace{0.1in} & 1.2, 0.2, 0.1 \\
trigger
           & 0.0, 0.0, 0.0 \hspace{0.1in} & 0.1, 0.0, 0.0 \hspace{0.1in} &
             0.1, 0.0, 0.0 \hspace{0.1in} & 0.0, 0.0, 0.0 \\
vertex & 0.0, 0.0, 0.1 \hspace{0.1in} & 0.1, 0.0, 0.0 \hspace{0.1in} &
             0.1, 0.0, 0.0 \hspace{0.1in} & 0.0, 0.0, 0.0 \\
tracking
           & 0.0, 0.0, 0.1 \hspace{0.1in} & 0.0, 0.0, 0.0 \hspace{0.1in} &
             0.0, 0.0, 0.0 \hspace{0.1in} & 0.0, 0.0, 0.0 \\
\br 
\end{tabular}
\end{indented}
\end{table}

\begin{table}[htb]
\caption{\label{TableIX}
Same as Table~\ref{TableVIII} except absolute values of errors (in percent) for Cu-Cu collisions at
$\sqrt{s_{NN}}$ = 200~GeV relative to the reference values in Table~\ref{TableVI} as explained in the text.}
\begin{indented}
\vspace{0.1in}
\item[]\begin{tabular}{@{}ccccc} \br 
Error & \multicolumn{4}{c}{{\bf Cu-Cu 200 GeV Errors}} \\
Source & $\Delta \langle b \rangle$(\%) & $\Delta \langle N_{part} \rangle$(\%) & $\Delta \langle N_{bin} \rangle$(\%) & $\Delta \nu$(\%) \\
\mr 
$c_{pt,m}$ & 1.0, 1.0, 1.0 \hspace{0.1in} & 0.2, 0.4, 0.5 \hspace{0.1in} &
             0.2, 1.1, 2.6 \hspace{0.1in} & 0.2, 0.8, 2.0 \\
$z_{pt,m}$ & 1.9, 1.8, 1.7 \hspace{0.1in} & 2.1, 1.8, 0.8 \hspace{0.1in} &
             3.2, 4.3, 3.2 \hspace{0.1in} & 1.2, 2.5, 2.3 \\
$\sigma_{inel}$ & 0.2, 0.2, 0.2 \hspace{0.1in} & 0.2, 0.4, 0.4 \hspace{0.1in} &
                  0.5, 1.5, 2.2 \hspace{0.1in} & 0.3, 1.1, 1.8 \\
$x$        & 0.0, 0.0, 0.0 \hspace{0.1in} & 0.0, 0.0, 0.0 \hspace{0.1in} &
             0.2, 0.0, 0.1 \hspace{0.1in} & 0.2, 0.1, 0.1 \\
$a$        & 0.1, 0.1, 0.3 \hspace{0.1in} & 0.8, 0.2, 0.1 \hspace{0.1in} &
             1.2, 0.4, 0.2 \hspace{0.1in} & 0.4, 0.2, 0.1 \\
$\cal P$-form & 0.3, 0.1, 1.9 \hspace{0.1in} & 2.8, 0.1, 0.5 \hspace{0.1in} &
             4.7, 0.5, 0.6 \hspace{0.1in} & 1.5, 0.5, 0.3 \\
trigger
           & 0.0, 0.0, 0.1 \hspace{0.1in} & 0.2, 0.1, 0.0 \hspace{0.1in} &
             0.2, 0.1, 0.0 \hspace{0.1in} & 0.1, 0.0, 0.0 \\
vertex & 0.0, 0.0, 0.0 \hspace{0.1in} & 0.0, 0.0, 0.0 \hspace{0.1in} &
             0.2, 0.0, 0.0 \hspace{0.1in} & 0.0, 0.0, 0.0 \\
tracking
           & 0.0, 0.0, 0.0 \hspace{0.1in} & 0.0, 0.0, 0.0 \hspace{0.1in} &
             0.0, 0.0, 0.0 \hspace{0.1in} & 0.0, 0.0, 0.0 \\
\br 
\end{tabular}
\end{indented}
\end{table}

\begin{table}[htb]
\caption{\label{TableX}
Percent changes in collision geometry bin average quantities relative to the
reference values in Tables~\ref{TableII} - \ref{TableVII}
due to background contamination
of the peripheral collision events as explained in Sec.~\ref{Sec:results}.
Background contamination errors
for centrality bins not listed were negligible.}
\begin{indented}
\vspace{0.1in}
\item[]\begin{tabular}{@{}cccccc} \br 
 & \multicolumn{5}{c}{{\bf Background Contamination Errors}} \\
System            & Centrality  &  $\Delta \langle b \rangle$ &
$\Delta \langle N_{part} \rangle$ & $\Delta \langle N_{bin} \rangle$
& $\Delta \nu$ \\
    &   (percent)  & (\%)  & (\%) & (\%) & (\%) \\
\mr 
Au-Au 200~GeV  & 90-100  & -0.3  &  5.6  &  7.9  &  1.8  \\
23\% contamination  & 80-90   &  0.2  &  -1.6  &  -1.7  &  -0.1  \\
\vspace{0.2cm}
in 88-100\% bin & & & & & \\
Au-Au 130~GeV  & 90-100  & -0.3  &  4.8  &  6.7  &  1.6  \\
22\% contamination  & 80-90   &  0.2  &  -1.5  &  -1.5  &  0.0  \\
\vspace{0.2cm}
in 91-100\% bin & & & & & \\
Au-Au 62~GeV  & 90-100  & -0.1  &  1.3  &  1.6  &  0.3  \\
18\% contamination  & 10-50$^{\rm a}$
                         &  -0.1  &  0.2  &  0.4  &  0.1  \\
\vspace{0.2cm}
in 96-100\% bin & & & & & \\
\hline
Au-Au 20~GeV$^{\rm b}$
              &  90-100  &  -0.1  &  2.7  &  3.5  &  0.8  \\
              &  80-90   &   0.0  &  0.0  &  0.1  &  0.0  \\
\vspace{0.2cm}
       & 40-70$^{\rm a}$       &  -0.1  &  0.5  &  0.7  &  0.2  \\
\vspace{0.2cm}
Cu-Cu 200~GeV$^{\rm b}$  &  90-100  &  -0.2  &  1.7  &  2.3  &  0.8  \\
\vspace{0.2cm}
Cu-Cu 62~GeV$^{\rm b}$   &  90-100  &  -0.1  &  1.5  &  2.0  &  0.6  \\
\hline \hline
\end{tabular}
\item[] $^{\rm a}$ Average percent changes within the combined centrality bin.
\item[] $^{\rm b}$ Reference uncertainties assuming 10\% background contamination in the number of collisions in the
nominal 90-100\% centrality bin, equivalent to 1\% overall background.
\end{indented}
\end{table}

\begin{table}[h]
\caption{\label{TableXI}
Upper half-max end-point positions $N_{part,max}$ and $N_{bin,max}$ from the
power-law distributions $d\sigma/d(N_{part}/2)^{1/4}$ and 
$d\sigma/dN_{bin}^{1/6}$, respectively, for the six collision systems studied here.}
\begin{indented}
\item[]\begin{tabular}{@{}ccc} \br 
System   &   $N_{part,max}$  &  $N_{bin,max}$  \\
\mr 
Au-Au 200 GeV   &   379.3   &   1166   \\
Au-Au 130 GeV   &   377.5   &   1082   \\
Au-Au 62 GeV    &   374.9   &    988   \\
Au-Au 20 GeV    &   373.4   &    926   \\
Cu-Cu 200 GeV   &   115.8   &    230   \\
Cu-Cu 62 GeV    &   112.7   &    196   \\
\br 
\end{tabular}
\end{indented}
\end{table}


\appendix
\section{ }

Multiplicity frequency distributions for high energy minimum-bias proton-proton
collisions are well described by a negative binomial distribution (NBD)~\cite{nbd,ua5} given by
\bea
{\cal P}_{\rm NBD}(n,\langle n \rangle,k) & = &
   \left( \begin{array}{c}
            n+k-1 \\
              k-1\\
          \end{array} \right)
\left( \frac{\langle n \rangle / k}{1 + \langle n \rangle / k} \right)^n
  (1 + \langle n \rangle / k)^{-k}.
\label{EqA1}
\eea
In Eq.~(\ref{EqA1}) $n$ and $\langle n \rangle$ are random and mean multiplicities, the
variance is $\langle n \rangle (1 + \langle n \rangle / k)$, the variance
excess relative to Poisson statistics is $\langle n \rangle^2/k$, and the
multiplicity fluctuation variance excess per final-state particle is
$\langle n \rangle / k$. The latter expression serves as an operational definition of
parameter $k$. For p-p collisions both $\langle n \rangle$ and $1/k$ increase
approximately linearly with ln($s$)~\cite{ua5} as expected from pQCD cross
sections for parton scattering.

Multiplicity fluctuation variance excess per final-state particle within a
given acceptance is equal to the integral of two-particle correlations on
relative momentum coordinates within the same acceptance~\cite{clt,delsign,ptscale}.
For Au-Au collisions at 130~GeV~\cite{axialci} and at 62 and 200~GeV~\cite{axialciQM08}
the integral of charged particle correlations for $|\eta| \leq 0.5$ and
$2\pi$ azimuth acceptance is dominated by a two-dimensional (2D) correlation
peak at small relative opening angles. The magnitude of this correlation
integral increases roughly with $(\nu - 1){\rm ln}\sqrt{s_{NN}}$.
Multiplicity fluctuation variance excess per final-state particle was therefore approximated by
\bea
\frac{\langle n \rangle}{k} \left[ \sqrt{s_{NN}},\nu \right]
& = &
\frac{\langle n \rangle}{k} \left[ \sqrt{s},{\rm p-p} \right]
\left[ 1 + \frac{(\kappa - 1)(\nu - 1)}{(\nu_{0-5\%} - 1)} \right]
\label{EqA2}
\eea
where quantity $\langle n \rangle / k [\sqrt{s},{\rm p-p}]$ is the variance
excess for p-p collisions, $\nu_{0-5\%}$ is the average number of binary
collisions per incident participant nucleon for the most-central (0-5\%)
A-A data, and $\kappa$ is taken to be the ratio of the 2D correlation peak
amplitude for most-central A-A collisions to that for p-p collisions.

Values of $\langle n \rangle / k$ for p-p collisions were obtained from
UA5~\cite{nbd,ua5} using either measured values of $k$ or the UA5 energy
dependent fit function for $k$. $\langle n \rangle$ for $|\eta| \leq 0.5$
is given by the values of $n_{pp}$ assumed in this analysis.  Simulations
demonstrated that the values of $k$ obtained by UA5 with very large $\eta$
acceptance remained approximately constant at the smaller $\eta$
acceptance assumed in this analysis. Values of quantities
$\langle n \rangle / k [\sqrt{s},{\rm p-p}]$ were 0.09, 0.24, 0.40 and 0.51
for $\sqrt{s}$ = 20, 62, 130 and 200~GeV, respectively.
$\kappa$ for Au-Au collisions at 20, 62, 130 and 200~GeV were estimated to
be 4.8, 5.0, 5.4 and 5.5, respectively, from analysis of correlation data
at 130~GeV~\cite{axialci} assuming ln$\sqrt{s}$ scaling. The remaining
values for Cu-Cu at 62 and 200~GeV were respectively 3.4 and 3.7
assuming $(\nu - 1)$ scaling.  Finally, for each simulated collision
with event-wise $\nu$ and $\bar{N}_{ch}$ from Eq.~(\ref{Eq1}) NBD distribution
Eq.~(\ref{EqA1}) was sampled to obtain the event-wise multiplicity.

\section{ }

Fit recovery of the nominal $dN_{evt}/dN_{ch}$ distribution following
shifts in the nuclear density parameters and $\sigma_{inel}$ was achieved
by constructing a perturbative correction, $\delta(d\bar{N}_{ch}/d\eta)$,
which was added to the right-hand side of Eq.~(\ref{Eq1}).
Parameter shifts affect the collision
distributions on $N_{part}$ and $N_{bin}$ and in turn the
$dN_{evt}/d{\bar N}_{ch}$ and $dN_{evt}/dN_{ch}$ distributions.
The fit recovery term is defined
such that the nominal $dN_{evt}/d{\bar N}_{ch}$ distribution is
maintained when the frequency distribution on $N_{part}$ changes.
Running integrals of the distributions $dN_{evt}/d{\bar N}_{ch}$ and
$dN_{evt}/d(N_{part}/2)$
produce a one-to-one correspondence between ${\bar N}_{ch}$ and $N_{part}/2$
as a function of the number of summed events.
Using the power-law representation this running
integral relation is given by,
\bea
\int_{{\bar N}_{ch,min}^{1/4}}^{{\bar N}_{ch}^{1/4}} d{\bar N}_{ch}^{\prime~1/4}
\frac{dN_{evt,D}}{d{\bar N}_{ch}^{\prime~1/4}}
    =
\int_1^{(N_{part}/2)^{1/4}} d(N^{\prime}_{part}/2)^{1/4}
\frac{dN_{evt,D^{\prime}}}{d(N^{\prime}_{part}/2)^{1/4}}
\label{Eq13}
\eea
which defines a locus of ordered pairs $[(N_{part}/2)^{1/4},{\bar N}_{ch}^{1/4}]$,
which can be compactly expressed as a discrete function defined by
\bea
{\bar N}_{ch}^{1/4}  &  =  & {\cal N}^{1/4}_{ch,D^{\prime} \rightarrow D}
[(N_{part}/2)^{1/4}].
\label{Eq14}
\eea
In Eqs.~(\ref{Eq13}) and (\ref{Eq14}) subscripts $D$ and $D^{\prime}$
refer to the nominal (reference) distribution when
$D=R$ and to the shifted distribution when $D=S$. Lower limit
${\bar N}_{ch,min}^{1/4} = n_{pp}^{1/4}$
for the reference distribution. In order to recover the nominal
$dN_{evt}/d{\bar N}_{ch}$ distribution the correction term must be computed
by the following difference,
\bea
\delta(\frac{d\bar{N}_{ch}}{d\eta}) & = & {\cal N}_{ch,S \rightarrow R}
[(\frac{N_{part}}{2})^{\frac{1}{4}}] - {\cal N}_{ch,S \rightarrow S} [(\frac{N_{part}}{2})^{\frac{1}{4}}].
\label{Eq15}
\eea
Throughout this fit recovery procedure the parameters of ${\cal P}(N_{ch},\bar{N}_{ch})$ in Eq.~(\ref{Eq2}) remained fixed.

Numerically stable results with one-million minimum-bias simulated collisions
required coarse binning of the Monte Carlo distributions
$dN_{evt,D}/d{\bar N}_{ch}^{1/4}$ and $dN_{evt,D}/d(N_{part}/2)^{1/4}$ prior
to evaluation of Eqs.~(\ref{Eq13}) - (\ref{Eq15}). When parameters $c_{pt,m}$ and
$\sigma_{inel}$ were shifted excellent fit recovery was achieved with centrality bins
corresponding to 0-5\%, 5-10\%, 10-40\%, 40-70\% and 70-100\% total cross section
fractions. When diffusivity parameter $z_{pt,m}$ was shifted the optimum
binning was 0-5\%, 5-10\%, 10-20\%, 20-40\%, 40-60\%, 60-80\% and 80-100\%.
The power-law representation in
Eq.~(\ref{Eq13}) exploits the approximately uniform statistics throughout
the domain of the integrands, making numerical stability of the results less problematic.

Variations in the density parameters and $\sigma_{inel}$ produced typical
changes in $\langle N_{ch} \rangle$ as a function of centrality of order several percent, up to 5\%.
Fit recovery in terms of $\langle N_{ch} \rangle$
was typically within a few tenths of a percent and
always less than about 1\% where the magnitude of
$\delta(d\bar{N}_{ch}/d\eta)$ relative
to $d\bar{N}_{ch}/d\eta$ was less than about 1\%, 4\% and 1.5\%
for changes in $c_{pt,m}$, $z_{pt,m}$ and $\sigma_{inel}$,
respectively, for all six collision systems.

A similar method was used to account for shifts in $dN_{evt}/dN_{ch}$
representing the systematic uncertainties in trigger inefficiency,
background contamination, and collision vertex reconstruction inefficiency.
For these simulations $dN_{evt}/d(N_{part}/2)^{1/4}$
remained fixed while $dN_{evt}/d{\bar N}_{ch}^{1/4}$ varied.
The required correction is given by,
\bea
\delta(\frac{d\bar{N}_{ch}}{d\eta}) & = & {\cal N}_{ch,R \rightarrow S}
[(\frac{N_{part}}{2})^{\frac{1}{4}}] - {\cal N}_{ch,R \rightarrow R} [(\frac{N_{part}}{2})^{\frac{1}{4}}].
\label{Eq16}
\eea

\section{ }

The Monte Carlo simulation model discussed in Sec.~\ref{Sec:discussion}
for transverse particle production and energy deposition in the STAR CTB for
p-p and Au-Au minimum bias collisions is described here.  For p-p collisions
the interaction point along the beam-line was randomly selected within
$\pm$25~cm of the geometrical center of the STAR detector~\cite{star}.  The
minimum-bias charged particle multiplicity was obtained by sampling the negative
binomial distribution (NBD) fitted to the UA5 $\sqrt{s}$ = 200~GeV p-p
data~\cite{nbd,ua5} with parameters $\langle n \rangle = 21.6$ and $k = 4.6$. The $p_t$,
$\eta$ and particle species ($\pi^{\pm}$, K$^{\pm}$, proton or antiproton) for
each charged particle produced in the collision was sampled from the
measured $dN_{ch}/dp_t$, $dN_{ch}/d\eta$,
and particle species distributions,
respectively.  Distribution $dN_{ch}/dp_t$ was assumed to be proportional to
$p_t \exp(-\beta m_t)$, where $m_t = \sqrt{p_t^2 + m^2}$, $m$ is the particle rest mass, and inverse
effective temperature $\beta$ = 5.7, 5.2 and 4.8~GeV$^{-1}$, for pions, kaons
and protons, respectively. The $\beta$ values were obtained by fitting
data~\cite{spectra2} within the
approximate range $m_t - m \leq 0.6$~GeV.
$p_t$ was restricted to be less than 2~GeV/$c$. Distribution $dN_{ch}/d\eta$
was obtained from the UA5~\cite{ua5} measurements for $\sqrt{s}$ = 200~GeV non-singly diffractive
p-p collisions for $|\eta| \leq 4.6$. The particle species probabilities
for pions, kaons and protons at low momentum were assumed to be 0.85, 0.085
and 0.065, respectively ~\cite{molnar,pid}.

Particle trajectories were extended outward from the collision point and
approximately transverse to the beam direction toward the STAR
Central Trigger Barrel (CTB) detector~\cite{startrig,star} which was
approximated by a uniform cylinder 220~cm in radius,
coaxial with the beam line, and having full $2\pi$ azimuthal coverage
and total longitudinal length of 484~cm located symmetrically about the
geometrical center of the STAR Time Projection Chamber~\cite{star}.
Charged particles were propagated in a 0.5~T solenoidal
magnetic field as helices until they either intersected or missed the CTB.
The simulated mean number of charged particles per unit pseudorapidity at $\eta = 0$
intersecting the CTB was consistent with the measured $n_{pp}$ from UA5~\cite{ua5}.

Energy deposition in the plastic scintillator material (CH) was
estimated by sampling the Landau distribution~\cite{landau} for the calculated
most-probable energy loss~\cite{pdg} for 1~cm thick plastic taking into account
the range of crossing angles
between the helical trajectories and the cylindrical detector.
The areal density (effective thickness) along the path through the material
was estimated to be
(1.032~gm/cm$^2)/\cos(\alpha)$~\cite{pdg} where $\alpha$
is the angle of incidence between the particle trajectory helix and a
normal to the detector cylinder at
the intersection point.  The mean electron excitation (ionization) energies
for carbon and hydrogen were assumed to be 80~eV and 20~eV,
respectively~\cite{pdg}.  Energy depositions in the carbon and hydrogen
components of the detector material were added together.  Light attentuation
in the plastic scintillator (380~cm attenuation length~\cite{startrig}) was included
in the simulated CTB output signal.  For each p-p collision the total output
from each produced charged particle which deposited energy in the CTB
scintillators was summed, resulting in a simulated result for the quantity
$SU\!M$ in Sec.~\ref{Sec:discussion}. The p-p results in Fig.~\ref{Fig4}
correspond to one-million triggered events with non-zero CTB signal. The
resulting trigger output frequency distribution, $dN_{trig}(pp)/dSU\!M$,
served as input for the Au-Au trigger simulations described next.

For Au-Au collisions the Monte Carlo Glauber model presented here was used to generate
an ensemble of $\sqrt{s_{NN}}$ = 200~GeV simulated minimum-bias collisions. For
each Au-Au collision the $dN_{trig}(pp)/dSU\!M$ distribution was sampled
$\bar{N}_{ch}/n_{pp}$ times corresponding to the average number of p-p
collisions required to generate mean multiplicity $\bar{N}_{ch}$.  The
sampled values of $SU\!M$ were added to obtain the total CTB trigger output
for each simulated Au-Au collision.  The results in Fig.~\ref{Fig4} correspond
to one-million Au-Au triggers with non-zero CTB output.

\end{document}